\documentclass[a4paper,fleqn,usenatbib]{mnras}

\usepackage{amsmath}	
\usepackage{times,txfonts,deluxetable,chngcntr}
\usepackage[T1]{fontenc}
\usepackage{ae,aecompl}

\usepackage{graphicx,caption}	
\usepackage{amssymb}	
\usepackage[english]{babel}
\usepackage{array}
\usepackage{multirow}
\usepackage{threeparttable}

\newcolumntype{M}[1]{>{\centering\arraybackslash}m{#1}}

\def\arcsecpoint{$''\!.$}
\def\degree{$^{\circ}$}

\title[NGC 4151]{Mass Outflow of the X-ray Emission Line Gas in NGC~4151}

\author[S. Kraemer et al.]{
{S.B. Kraemer $^{1}$, T.J. Turner $^{2}$, J.D. Couto$^{3}$, D.M. Crenshaw $^{4}$, H.R. Schmitt $^{5}$,}
\newauthor{ M. Revalski $^{6}$, and T.C. Fischer $^{6}$}    
\\
$^1$Department of Physics, Institute for Astrophysics and Computational Sciences, The Catholic University of America,\\
 Washington, DC 20064, USA; kraemer@cua.edu\\
$^2$Department of Physics, University of Maryland Baltimore County, 
   Baltimore, MD 21250 U.S.A\\
$^3$Department of Physics and Astronomy, Johns Hopkins University, Baltimore, MD 21210, USA\\
$^4$Department of Physics and Astronomy, Georgia State University, 25 Park Place, Room 631, Atlanta, GA 30303, USA\\
$^5$Naval Research Laboratory, Washington, DC 20375, USA\\
$^6$Space Telescope Science Institute, Baltimore, MD 21218, USA\\}

\date{Accepted XXX. Received YYY; in original form ZZZ}

\pubyear{2019}

\begin{document}
\label{firstpage}
\pagerange{\pageref{firstpage}--\pageref{lastpage}}
\maketitle

\begin{abstract}
We have analysed {\it Chandra}/High Energy Transmission Gratings spectra of the X-ray emission line gas in the Seyfert galaxy NGC~4151. The zeroth order spectral images show extended H- and He-like O and Ne, up to a distance $r \sim$ 200 pc from the nucleus. Using the 1st order spectra, we measure an average line velocity $\sim -230$ km s$^{-1}$, suggesting significant outflow of X-ray gas. We generated Cloudy photoionisation models to fit the 1st order spectra; the fit required three distinct emission-line components. To estimate the total mass of ionised gas and the mass outflow rates, we applied the model parameters to fit the zeroth order emission-line profiles of Ne~IX and Ne~X. We determined the total mass of $\approx 5.4 \times$ 10$^{5}$ M$_\odot$. Assuming the same kinematic profile as that for the [O~III] gas, derived from our analysis of {\it Hubble Space Telescope}/Space Telescope Imaging Spectrograph spectra, the peak X-ray mass outflow rate was $\approx 1.8$ M$_\odot$ yr$^{-1}$, at $r \sim  150$ pc. The total mass and mass outflow rates are similar to those determined using [O~III], implying that the X-ray gas is a major outflow component. However, unlike the optical outflows, the X-ray outflow rate does not drop off at $r >$ 100 pc, which suggests that it may have a greater impact on the host galaxy.

\end{abstract}




\section{Introduction}\label{sec:Intro}

\subsection{General Background}\label{sec:GenB}

It is believed that Active Galactic Nuclei (AGN) are powered by mass accretion onto a supermassive black hole (SMBH) at the gravitational centre of the host galaxy \citep[e.g.,][]{rees87a}. The SMBH and surrounding accretion disk can generate  relativistic jets and powerful winds. AGN-driven mass outflows can effect conditions in the host galaxy, in a process referred to as ``AGN feedback'' \citep{begelman04a}. Once the AGN is activated, jets or winds evacuate the bulge of the host galaxy, quenching star-formation. This ultimately produces the 
observed relationship between the mass of the SMBH and the bulge, as evidenced
by the $M_{\rm bh}-\sigma$ relationship \citep[e.g.,][]{gebhardt00a}. 

Although radio jets can be powerful enough to drive feedback, only $\sim 15\%$ of radiation-dominated AGN are radio-loud \citep{netzer15a}.  On the other hand, the fact that there is evidence that AGN-driven winds are present in $\gtrsim 50\%$ of AGN \citep{ganguly08a} suggests that they can span a much larger solid angle and, therefore, could be the principle feedback mode. Such winds have been detected along our line-of-sight to AGN in the form of blue-shifted absorption lines in high-resolution UV spectra \citep{crenshaw99a} and X-ray spectra \citep{kaastra00a, kaspi01a}. Outflow velocities are typically on the order of 100s to a few 1000s km s$^{-1}$, but can be as large as $0.1c-0.5c$, e.g., in Broad Absorption Line QSOs \citep[e.g.,][]{turnshek84a} and in AGN that show highly-blueshifted X-ray absorption, i.e., ``Ultra Fast Outflows'' \citep[e.g.,][]{tombesi11a}. Nevertheless, in most cases the radial distances and global covering factors of the absorbers are not well-constrained, hence the total masses and mass outflow rates and, therefore, the impact on the host galaxies, are not easily determined.   

On the other hand, there has been extensive study of mass outflow in emission in AGN in the local Universe ($z < 0.1$) \citep[e.g.,][]{fischer13a}. The targets are typically Seyfert galaxies, which are AGN with modest
luminosities, $L_{bol} \leq 10^{45}$ erg s$^{-1}$, although, more recently,
\citet{fischer18a} have extended the study to QSO2s, which are up to several orders of magnitude more luminous. 

While the details of AGN classification have been reviewed numerous times \citep[e.g.,][]{peterson97a, netzer15a}, it is useful to briefly summarise them. Generally, AGN can be classified via their optical emission. Type 1s
possess broad permitted lines, with full widths at half maximum $>$ a few 1000
km $^{-1}$, narrower forbidden lines, with FWHM $\leq 1000$ km s$^{-1}$ and non-stellar continua. Type 2s have narrow permitted and forbidden lines, and their continua are typically dominated by starlight from the host galaxy. The optical continuum is thought to be thermal emission from the assumed accretion disk. The broad lines are formed in dense gas (n$_{H} > 10^{8}$ cm$^{-3}$) in a region
from several to 100 light days from the central SMBH, referred to as the broad line region (BLR). The narrow components of the permitted lines and the forbidden lines are formed in a region extended from several pcs to several kpcs, called the narrow line region (NLR). The so-called  unified model for AGN \citep{antonucci93a} postulates that both types are
intrinsically the same, but the BLR and continuum source in Type 2s are obscured
by a dense, molecular torus along our line-of-sight. 
The BLR gas is thought to have components from the accretion disk and AGN-driven winds
\citep[e.g.,][]{collin06a}, although the winds may include a strong rotational component, if they originate in the disk \citep[e.g.,][]{kashi13a}.  The NLR kinematics are consistent with combination of radial outflows and rotation \citep[e.g.,][]{fischer17a}. 

NLR-scale outflows in both Type 1 and 2 AGN  have been studied via {\it Hubble Space Telescope (HST)}/Space Telescope Imaging Spectrograph (STIS) observations \citep[e.g.,][]{fischer13a, fischer18a} and ground-based spectra, such as those obtained with the {\it Gemini}/Near Infrared Integral Field Spectrograph \citep[e.g.,][]{storchi10a, riffel13a, fischer17a}. The NLR outflows extend to kpc scales \citep{fischer18a}, where much of the nuclear star-formation occurs. Furthermore, most of the mass in these outflows is accelerated in situ, due to the radiation pressure from the AGN accelerating material in the disk of the host galaxy. Hence, the NLR-scale 
outflows directly reveal the interaction of the AGN with the host galaxy. 

Although the NLR outflows are often massive (e.g., mass of $\sim 10^{6}$ M$_\odot$, \citealt{collins09a, revalski18b}), and may inject
large amounts of kinetic energy and momentum into the interstellar medium (ISM) of the host galaxy \citep[e.g.,][]{revalski18a}, they do not appear to be able to escape the inner kpc of the host galaxy bulge 
\citep{fischer18a}.  This calls into question the effectiveness of AGN-driven outflows in feedback. However, these results are generally derived from optical or near-IR emission line
kinematics, they do not take into account the role of higher-ionisation gas.

Low-resolution X-ray spectra of Type 1 AGN, such as those obtained with the {\it Advanced Satellite for Cosmology and Astrophysics (ASCA)} revealed absorption by ionised gas along our line of sight
to the continuum source \citep{reynolds97a}, as well as ``spectral complexity'' below $<$ 1 keV \citep{george98a}, where the observed continuum emission was weak, due to absorption. 
Similar soft X-ray properties were detected in {\it ASCA} spectra of Type 2 AGN \citep{turner97a}. 
\citet{netzer93a} suggested that the absorber could be a source of the soft X-ray emission and that the spectral complexity would be resolved into emission lines
by more sensitive and higher resolution X-ray spectrographs. This prediction was confirmed in observations made with XMM-{\it Newton} \citep{sako00a, kinkhabwala02a, turner03a}
and {\it Chandra} \citep{ogle00a}. In general, the soft X-ray spectra showed emission lines from H- and He-like ions of C, N, O, Ne, Mg, Si and S, 
as well as radiative recombination continua (RRC) from those ions. 
          
{\it Chandra} X-ray imaging of the Seyfert 2 galaxy NGC~1068 showed that the X-ray emission line gas was extended and coincident with the optical NLR \citep{young01a}.
\citet{bianchi06a} determined that this was the case for most Seyfert galaxies. They further suggested that the ionisation mechanism, specifically
photoionisation by the AGN, was the same for both the X-ray and optical components. \citet{kallman14a} performed a detailed photoionisation analysis of {\it Chandra}/High Energy Transmission Grating (HETG, \citealt{canizares05a}) spectra of NGC~1068. They estimated the total mass and mass outflow rates of the emission-line gas to be $M \approx 3.7 \times 10^{5}$ M$_\odot$ and $\dot{M}_{out} \sim 0.3$ M$_\odot$ yr$^{-1}$, respectively. \citet{bogdan17a} analysed {\it Chandra}/HETG spectra of the Seyfert 2 galaxy Mrk 3 and derived some constraints on the emission-line gas.
Although there have been efforts to model the X-ray NLR using {\it Chandra} imaging \citep[e.g.,][]{bianchi10a, gonzalez10a, maksym19a}, there have not been any other attempts to generate detailed physical models using HETG spectra.

\subsection{Previous Study of NGC 4151}\label{sec:4151Back}

The Seyfert 1.5 galaxy, NGC~4151 ($z=0.0033$; \citealt{devaucouleurs91a}) has been observed extensively, across most of
the electro-magnetic spectrum. From reverberation mapping, \citet{bentz15a} derived a black hole mass of  log M$_\odot=7.555^{+0.051}_{-0.047}$. 
The most recent distance estimate is $\approx 15$ Mpc (M.M. Fausnaugh et al., in preparation), which gives a scale of 70 parsecs per arcsecond.  Most relevant to the work presented here are the results of spatially-resolved
spectroscopy, obtained with {\it HST}/STIS. Using STIS G430M slitless spectra, \citet{hutchings98a} were the first
to unequivocally show that the NLR gas kinematics were consistent with a bi-conical outflow. \citet{kraemer00b} performed a photoionisation analysis 
of NLR spectra obtained using the STIS G140L, G230L, G430L, and G750L gratings, through the 52$^{''}$X0\arcsecpoint1 aperture. They determined that the NLR gas is
photoionised by the EUV-X-ray radiation from the AGN. Furthermore, they argued that the gas is inhomogeneous, consisting of co-located radiation and matter
bounded components, of different ionisation states, and that the gas density decreases with radial distance.

As initially revealed in {\it Chandra}/HETG spectra, NGC~4151 possesses soft X-ray emission extending $\sim 1.6$ kpc from the nucleus \citep{ogle00a}. The X-ray emission  comprises  numerous emission-lines from H- and He-line ions of C, N, O, Ne, Mg, and Si. \citet{ogle00a} suggest that, while most of the emission-lines are consistent with an origin in a photoionised plasma, there is evidence for collisional enhancement of the Ly$\alpha$ lines of O~VIII, Ne~X, and Mg~XII.
However, it has been shown that these lines can be strengthened via photo-excitation \citep[e.g.,][]{sako00a}, hence there is no clear signature for a collisional component.

We had previously modelled the soft X-ray emission lines in NGC~4151 using archival XMM-{\it Newton}/Reflection Grating Spectrometer observations \citep{armentrout07a}.
We measured blue-shifted radial velocities for H- and He-like lines of C, N, O and Ne, with an average value of $\sim -250$ km s$^{-1}$.
The RRC were consistent with an origin in photoionised gas \citep{liedahl96a}. The He-like triplets indicated some enhancement of the resonance $1s2p$~$^{1}$P$_{1}$ $\rightarrow$
$1s^{2}$~$^{1}$S$_{0}$ lines (r-lines), compared to the intercombination $1s2p$~$^{3}$P$_{2}$ $\rightarrow$ $1s^{2}$~$^{1}$S$_{0}$, $1s2p$~$^{3}$P$_{1}$ $\rightarrow$
$1s^{2}$~$^{1}$S$_{0}$, and forbidden $1s2s$~$^{3}$S$_{1}$ $\rightarrow$ $1s^{2}$~$^{1}$S$_{0}$ lines (i- and f-lines). We attributed this to photo-excitation of the r-line \citep[e.g.,][]{sako00a}. We fit the emission line ratios and luminosities
with a 3-component photoionisation model, generated with the code Cloudy (for the current version, see \citealt{ferland17a}). The model values for ionisation parameter, log$U$\footnote{$U = \frac{Q}{4 \pi n_{\rm H} r^{2} c}$,
where, $n_{\rm H}$ is the hydrogen number density, $r$ is the distance to the ionizing source, $c$ is the speed of light, and the ionising luminosity, $Q = \int_{\nu_{o}}^{\infty} L_{\nu} d\nu$, where h$\nu_{o} = 13.6$eV.}, and column density, log($N_{\rm H}/cm^{2}$) were:
$-0.5$, 20.5; 0.0, 23; and, 1.3, 23. We assumed that the emission-line gas was associated with the UV and X-ray absorbers discussed in \citet{kraemer05a, kraemer06a}, therefore, the radial
distances were all $< 10^{18}$ cm. Although the models were not physically inconsistent, the RGS data do not posses the spatial 
resolution to reveal the extent of the emission line gas, hence our analysis did not provide much insight into the mass of ionised gas or mass outflow rates.
  
Based on their analysis of a 200 ks {\it Chandra}/ACIS image of NGC 4151,  
\citet{wang11a, wang11b, wang11c} performed a detailed study of the X-ray emission. In agreement with \citet{ogle00a}, they found X-ray emission in the 0.3 - 1 keV (``soft'') band, extending 30 pc to 1.3 kpc, in a northeast (NE) to southwest (SW) direction. The emission cuts off in the NE, at a point coincident with a CO gas lane in the host galaxy. On the other hand, the 1 - 7 keV emission appeared unresolved. The soft emission within 2$^{''}$ of the nucleus shows a linear structure, and there are discrete knots within 1$^{''}$. By isolating regions of the spectrum, they were able to map the O~VII, O~VIII, and Ne~IX emission, which appeared spatially correlated with the optical [O~III] $\lambda$5007 emission detected by {\it HST}/WFPC2 (see \citealt{kraemer08a}). They assumed that the [O~III] and X-ray emission arise from a single, heterogeneous (in density) photoionised medium. Using the [O~III] velocities presented in \citet{crenshaw15a}, and a density law $n_{\rm H} \propto r^{-2}$, where $n_{\rm H}$ is the hydrogen number density and $r$ is the distance to the ionising source,
\citet{wang11c} obtained a mass outflow rate of $\dot{M}_{out} \approx$ 2 M$_\odot$ yr$^{-1}$, at $r = 130$pc, and a peak Kinetic Luminosity $L_{KE} \approx 1.7 \times 10^{41}$ erg s$^{-1}$, for the X-ray gas. For comparison, \citet{crenshaw15a} derived a peak value of $\dot{M}_{out} \sim 3$ M$_\odot$ yr$^{-1}$ at 70 pc, and a peak $L_{KE} \approx 4.3 \times 10^{41}$ erg s$^{-1}$ at 90 pc, for the optical gas.  

In the present paper, we will use {\it Chandra}/HETG spectra to determine the physical conditions within the X-ray emission-line gas.
In Section \ref{sec:Obs}, we present the method and results of the spectral and imaging analysis. In Section \ref{sec:Modelling}, we present the 
photoionisation analysis of the $\pm$1st order spectra. In Section \ref{sec:ExtEmm} we apply the modelling results to the analysis of the extended
emission measured using the zero-th order spectra. Finally, in \ref{sec:DISC}, we discuss the dynamics and origin of the X-ray outflow.
 
\section{Chandra Observations}\label{sec:Obs}

This paper presents imaging and spectroscopic results from our {\it Chandra} observations of NGC~4151 
at two epochs during 2014, first presented in \citet{couto16a}.  The observations total 241 ks in exposure time. These data were split into two epochs due to constraints in the 
 roll angle alignment to meet the observation goals, i.e., to allow us to have the cross-dispersion
direction correctly oriented to observe the  extended emission of NGC 4151 (see \citealt{ogle00a}).
  
 The first dataset, OBSID16089, was an exposure of 172 ks, obtained over the period 2014 February 12 - 14 while 
the second, OBSID 16090, was an exposure of 69 ks taken  2014 March 08 - 09. Both observations utilised the HETG  
grating assembly 
dispersed onto the S-array of the Advanced CCD Imaging
Spectrometer (ACIS). CCDs S1 to S5 were used.  The HETG 
 consists of a pair of grating arrays. The High Energy Grating (HEG) covers $\sim 0.8 - 8.5$ keV,  16-1.5 {\AA} with a  spectral resolution of  0.012 {\AA} FWHM.  The Medium Energy Grating (MEG) covers 31-2.5 {\AA} with spectral resolution 0.023 {\AA} FWHM. 
 
Standard level 2 events files were taken as the starting point and   {\it Chandra} Interactive Analysis of Observations (CIAO)  software version 4.10 together 
with {\sc CALDB} 4.7.9 and HEAsoft v6.24  were used for the data analysis. 
The mean source flux for the 2014 observations was $F_{2-10} = 6.3 \times 10^{-11}$ erg cm$^{-2}$ s$^{-1}$. 
The total number of counts in the summed (positive and negative) first order HEG spectra 
were 35938 counts, with a corresponding count rate  $0.1483\pm0.0010$ count s$^{-1}$ and, for MEG, 
32115 counts, with a count rate of $0.1326\pm0.0009$ count s$^{-1}$. 
The background level was negligible compared to the source.
The full, unfolded spectrum is shown in Figure \ref{fig:FullSpec}.

\subsection{Imaging Analysis}\label{sec:Images}

Based upon previous spectral and imaging analysis of NGC~4151 
\citep[e.g.,][]{ogle00a, wang11a}, we selected twelve wavebands that allowed us to study the spatial dependence at wavelengths dominated by emission lines of interest. We did not subtract the continuum from the images. Two line-free continuum bands were also chosen for comparison with the emission line maps.  

The two OBSIDs were combined for imaging analysis, using {\sc CIAO} task {\verb+merge_obs+}. This task creates an exposure map for each observation that accounts for the dither pattern and allows correct summation of observations with different roll angles. This yielded images in units of photons cm$^{-2}$ s$^{-1}$. The images created were sampled at the maximum spatial resolution of ACIS, i.e, with pixels of $\sim 0.5''$ on a side. These images were smoothed with a Gaussian having a 1$''$ kernel radius. The colour maps were adjusted to show the same flux scale for each image.  

As shown in Figure \ref{fig:Zeroth}, the continuum, both at 2.05-2.25 keV and 5.60-6.00 keV, is roughly symmetric and centred on the nucleus. The
higher flux in the higher energy bin is the result of lower absorption than at lower energies (see also Figure \ref{fig:FullSpec}). There is some evidence for weak emission
in the 2.05-2.25 keV band, towards the NE, but its origin is uncertain. The Si~K$\alpha$ and Fe~K$\alpha$ lines are also
strongly centrally peaked, again with a slight extension to the NE. However, the location of the peak of the emission encompasses
the innermost region of the AGN, hence most of the K line emission is consistent
an origin in the putative torus. As such, we have modelled it as due to reflection (see below). 

The lines with energies $>$ 1.3 keV, e.g., Si~XIV, Si~XIII, and Mg~XII, are all quite compact, in agreement with \citet{wang11a} regarding
the morphology of the emission in the 1 - 7 keV band. On the other hand, there is evidence that Mg~XI is extended along the NE-SW direction.
Lines with energies $\leq$ 1.05 keV, i.e., Ne~X, Ne~IX, O~VIII, and O~VII, are all clearly extended along a NE-SW direction, out to an projected distance of $>$ 200 pc.
Although this is smaller than that noted by \citet{wang11a, wang11b}, it is not unexpected considering the lower effective area of the HETG and
the loss in ACIS sensitivity at energies $<$ 1 keV \citep{odell13a}.  We revisit the nature of this extended emission in Section \ref{sec:ExtEmm}.

\subsection{Global Spectral Analysis}\label{sec:Spectra}

We first analysed spectra over the entire NLR, from which we obtained emission line and RRC fluxes and line velocity centroids. 
Spectra were extracted using the default extraction criteria and were pulled from the {\sc pha2} files provided with the 
pipeline products for each OBSID. ARF and RMF files were generated for each OBSID, then the pairs of positive and negative 
spectra were combined for the first order spectra.  The data were binned to a minimum of 10 counts per spectral bin such that the $\chi^{2}$ statistic could be utilised
\citep[e.g.,][]{arzner07a}. However, in order to determine whether our results were biased by our binning limit,
we also fit the data binned to a minimum of 20 counts per spectral bin. The results of the test are discussed in Appendix \ref{sec:App}. In summary,
there were no significant differences, except that the coarser binning washed out several
emission features. Hence, we have opted to use the 10 count minimum for our analysis. 
The fits were performed with the positive and negative parts of the spectra kept separate. The two OBSIDs were fitted jointly, rather than being co-added. 
For the spectral plots we combined the positive and negative orders, with both OBSIDs co-added, and applied additional binning, for clarity in the figures. 

The hard energy band of the spectrum is dominated by nuclear continuum emission, and the prominent Fe K$\alpha$ line is evident, at 6.4 keV ($\lambda = 1.94$ \AA). At
energies $<$ 2 keV, there are strong emission-lines from H- and He-like O, Ne, Mg, and Si,
and there is a Si~K$\alpha$ line, at 1.74 keV ($\lambda = 7.12$ \AA).

\subsubsection{Emission Line and RRC measurements}\label{sec:Lines}

For the lines, we fitted a Gaussian model and local continuum (using a power-law, with an un-constrained slope) for a restricted bandpass close to the line, using $\chi^{2}$ statistics.  The width of the bandpass used depended on the particular region, i.e. whether it was crowded or the line was isolated.  The line flux and observed energy were then fitted for most lines. In most cases, the line widths were frozen at 2 eV, because the lines were generally too weak to determine width. 
The RRCs were fitted using the {\sc XSPEC} model ``redge'', with local continuum fitted as a simple power-law, as for the lines. The RRC energies and widths were fixed. Subsequently, we fit for the RRC fluxes.

The resulting emission-line and RRC fluxes are listed in Table \ref{tab:Lines}. For those lines which were
detected both in the HETG and {\it XMM}/RGS spectra \citep{armentrout07a}, i.e., Ne~IX rif, Ne~X $\alpha$, O~VII rif,
and O~VIII $\alpha$, the measured fluxes agree, within the uncertainties. We were not able to obtain accurate widths, or corresponding temperatures, for
any of the RRCs. However, as we will show in Section \ref{sec:InitRes}, the photoionisation model predictions for the O~VII and O~VIII RRCs are
consistent with the data.

Due to the structure of the underlying continuum, the statistical significance of the emission features must be evaluated in
the context of the overall fitting of the data. We will address this in \ref{sec:Inputs}.  

\subsubsection{Emission Line Velocities}\label{sec:LineVels}

In Table \ref{tab:Lines}, we also list the radial velocities for the strongest emission lines. In Figure \ref{fig:Velocities}, we plot the velocities versus ionisation potential (IP);
given the uncertainties in the velocities for Ne~IX f, O~VII f, and O~VIII $\alpha$, we used the values in \citet{armentrout07a},
as well as their velocities for N~VII $\alpha$ and N~VI f, which were not detected in the HETG spectra. The average radial velocity is $v_{r} = -230^{-90}_{-370}$ km s$^{-1}$. 
Although there is evidence for a relationship between velocities and the ionisation potential for optical and UV lines in some Seyfert galaxies \citep{kraemer00a, kraemer09a},
which suggests a multi-component NLR, we found no strong evidence of such an relationship in NGC~4151 \citep{kraemer00b}. Therefore, while 
the X-ray emission line gas consists of components of different ionisation, as we show in Section \ref{sec:InitRes}, there is no evidence for kinematic differences
among them, which is consistent with the optical and UV result.

The X-ray emission lines velocities were measured using spectra covering the entire region detected in these observations.
Hence, they are a flux-weighted average. For comparison, \citet{crenshaw10a} derived an integrated [O~III] emission-line profile by
collapsing {\it HST}/STIS G430M, which spanned most of the NLR. They obtained $v_{r} = -60~(\pm 6)$~km~s$^{-1}$. However, their measurements included
red-shifted emission on the NE side of the nucleus. The emission line gas in that region lies behind the disk of the host galaxy \citep{das05a} and
could be covered by X-ray absorption similar to that described by \citet{couto16a}. In fact, soft X-ray emission lines can be effectively suppressed by relatively small
column density absorbers (e.g., $N_{\rm H} \leq 10^{21}$ cm$^{-2}$, \citealt{kraemer11a}). If the difference in $v_{r}$ is due to suppression of the
X-ray emission in the NE, it is highly plausible that the X-ray gas kinematics
are similar to that of the [O~III] gas. We revisit this in Section \ref{sec:MMdot2}.

\section{Modelling Analysis}\label{sec:Modelling}

\subsection{Inputs to the Models}\label{sec:Inputs}

Our principle goal for this analysis is to determine the physical properties of the global, i.e. unresolved, emission. 
To that end, we generated photoionisation model grids, computed using Cloudy (version 13.00; \citealt{ferland13a}\footnote{The models were rerun with the C17 version of Cloudy \citep{ferland17a}, with no significant differences in the model results.}).  We assumed a continuum source with the same spectral energy distribution
(SED) used in our analysis of the X-ray absorbers \citep{couto16a}, in the form of a power law $L{_\nu} \propto {\nu^{-\alpha}}$, with 
$\alpha$ = 1.0, for 1 $\times 10^{-4}$ eV $<$ h$\nu$ $<$ 13.6 eV, $\alpha = 1.3$, for 13.6 eV $<$ h$\nu$ $<$ 500 eV,
and $\alpha= 0.5$, for 500 eV $<$ h$\nu$ $<$ 30 keV,  with exponential cutoffs above and below the limits. In addition to the SED, the model results depend on the choice of input parameters, specifically: the radial distances of the emission-line gas with respect to the
central source ($r$), number density ($n_{\rm H}$),  and column density ($N_{\rm H}$) of the gas. The models are parameterised in terms of the ionisation parameter, $U$.

The continuum emission in NGC~4151 is highly variable, from optical to X-ray energies \citep[e.g.,][]{edelson96a}. For our models, we scaled the ionising luminosity, $Q$, using a rough average
UV flux at 1350 \AA~ observed over the last $\sim$ 20 yrs (see Figure 3 in \citealt{couto16a}), and computed $Q$ using our assumed SED and a distance of
15~Mpc. We obtained a value $Q = 3 \times 10^{53}$ photons s$^{-1}$. For comparison, using the same SED, \citet{wang11b} determined a value of $Q = 7.2 
\times 10^{53}$ photons s$^{-1}$, while \citet{couto16a} used $Q = 0.4 - 1.1 \times 10^{53}$ photons s$^{-1}$ to model the intrinsic absorption.    
Based on this continuum, the bolometric luminosity $L_{bol} = 1.4 \times 10^{44}$ erg s$^{-1}$ or 0.031 of the Eddington Luminosity.  
 
The elemental abundances in NLR emission-line gas in NGC~4151 were initially determined in our photoionisation modeling of STIS long slit spectra \citep{kraemer00b}.
Based on more recent estimates of solar elemental abundances \citep[e.g.,][]{asplund05a}, the abundances we had determined for NGC~4151 correspond to roughly 1.5 times solar.
We assumed these abundances in our most recent warm absorber study of NGC~4151 \citep{couto16a}, and have opted to assume the same values for the 
present paper,
as follows (in logarithm, relative to H, by number):
He: $-1.00$, C: $-3.47$, N: $-3.92$, O: $-3.17$, Ne: $-3.96$,
Na; $-5.76$, Mg: $-4.48$,  Al: $-5.55$, 
Si: $-4.51$,  P: $-6.59$, S: $-4.82$,  Ar: $-5.60$, 
Ca: $-5.66$,  Fe: $-4.4$, and Ni: $-5.78$.

We generated grids of photoionisation models over a range of values of U and $N_H$, using the default
energy resolution. We converted the Cloudy output to fittable 
grids using the \citet{porter06a} CLOUDY-to-{\sc XSPEC} interface, which maps the CLOUDY output onto a FITS format grid 
of either the additive emission components ({\sc ATABLES}) or multiplicative absorption coefficients 
({\sc MTABLES}). {\sc XSPEC} \citep{arnaud96a} works by interpolating between grid values in the fitting process.

The X-ray spectrum of NGC~4151 is complex, comprising several components: 1) heavily absorbed power-law continuum emission from the AGN, 2) reflection, including the Fe~K and Si~K fluorescence lines,
3) emission lines and RRCs, and 4) scattered or unabsorbed power-law continuum.  The counts in the data are dominated by the first two of these (see Figure \ref{fig:FullSpec}).  
Therefore, our approach was to obtain a satisfactory fit to the absorbed power-law and reflection components.
Next, we added {\sc ATABLES} for the emission features, other than those associated with reflection, e.g., the Fe~K and Si~K lines.
Since the emission features are weak compared to the continuum, the addition of the {\sc ATABLES} had a negligible effect
on the fit statistics. Nevertheless, once we had a good fit to the underlying continuum, we were able to test
for the statistical significance of the emission features, as we describe below. 
Then, we adjusted the parameters of the {\sc ATABLES} to optimise the fit to the statistically significant emission lines and RRCs. In the process,
we froze selected parameters from the earlier iterations; the final model parameters are listed in Table \ref{tab:Fits}.

In the analysis, we used data over 0.9-7.8 keV (13.78 \AA-1.59 \AA), for the HEG, and 0.5-6.0 keV (24.8 \AA-2.07 \AA), for the MEG.
To model the low ionization reflection 
we used the {\sc xillverEc} model, version 1.2.0 \citep{garcia13a, dauser13a}.
All models included the Galactic line-of-sight absorption,
N$_{\rm H,Gal} = 2.0 \times 10^{20}{\rm cm}^{-2}$ \citep{dickey90a}, parameterised using {\sc tbabs}.

The {\sc ATABLE} grid spanned a range from 22 $\leq$ log$N_{\rm H}$ $\leq 23$ and
$-0.5 \leq $ log$U$ $\leq 1.3$, in intervals of 0.1 dex; the ranges are consistent with those used
in \citet{armentrout07a}. We also assumed a turbulence velocity of 100 km s$^{-1}$, in agreement
with our previous studies \citep{armentrout07a, couto16a}. The {\sc MTABLES} were generated via a two-step process. First,
we generated a single zone, with parameters, log$U = 0.42$ and log$N_{\rm H} = 22.7$, to match
the high-ionization absorber, XHIGH, described in \citet{couto16a}. Then, using the continuum transmitted through
XHIGH, we generated an {\sc MTABLE} grid spanning a range from 22.4 $\leq$ log$N_{\rm H}$ $\leq 22.8$ and
$-1.8 \leq $ log$U$ $\leq 0.18$, in intervals of 0.1 dex. The rationale for the filtering of the
continuum is explained in \citet{kraemer05a}.   

The baseline model comprised a power-law continuum, whose slope reached a preferred value $\Gamma=1.50 \pm0.04$, consistent
with the SED used for the Cloudy models.  This power-law was absorbed by two zones ({\sc MTABLES}), the first with 100\% covering of the continuum
source and the second covering the first zone. Inclusion of the two zones improved the fit from
a reduced $\chi^{2}$ $=$ 5.92 for 3054 degrees of freedom (dof), to 1.26, for 3053 dof. The addition of the 
{\sc xillverEc} model improved the fit to a reduced $\chi^{2} = 1.03$ for 3053 dof.
While the inclusion an unabsorbed power-law, at 3\% of the flux of the continuum source, made a small statistical change ($\Delta \chi^{2} = 158$), visual inspection showed 
improvement of the continuum fitting, consistent with the suggestion by \citet{kraemer05a}. The overall continuum fit was optimised to isolate the emission features, hence, while roughly consistent with \citet{couto16a},
we did not require it to be an exact match.    

Having obtained a good fit to the absorbed power-law plus reflection, we were now able to assess the significance
of the emission features. To do so, we executed a sliding Gaussian test (see \citealt{turner03a}). Specifically, for each detector, we stepped a Gaussian profile across
the spectra in 0.1 \AA~ intervals. The Gaussian width, expressed as $\sigma$, was always fixed at 2 eV. The line was set to the first wavelength and fixed, 
then fit for flux and a $\Delta \chi^{2}$, showing the improvement to the fit for a narrow Gaussian being added at that energy, is returned. Then, the Gaussian was moved one step to next wavelength, and the process was repeated.
Thus, for each fit step, the line wavelength and width are fixed and the data were tested for the presence of a significant narrow line.
Note, when this test encounters a broad feature, e.g., an RRC or an unresolved He-like triplet, it returns a broad hump in $\Delta \chi^{2}$, as the Gaussian first encompasses the wings of the feature and, then, the peak, as it sweeps over it.

The results of the sliding Gaussian are shown in Figure \ref{fig:Slide}. Note, we did not extend the test to wavelengths $<$ 5.5\AA~, since we did
not detect strong emission lines in that region other than Fe~K. The emission features listed in Table \ref{tab:Lines} are all present.
Those emission features used to constrain the models were those that returned values of $\Delta \chi^{2}$ of more than -25, although we list the
those with weaker statistical significance to compare to the Cloudy predictions (see below). We did not test to wavelengths $>$ 15 \AA~, since the signal-to-noise of the
spectra were too low in that region for this method to return accurate results. However, as shown in Figure \ref{fig:Zeroth}, O~VIII and the O~VII rif lines
are clearly present in the spectra, hence we included them in our models.
We next added the {\sc ATABLES} to fit the detected emission features.   

As discussed in Section \ref{sec:Obs}, there are emission lines over a wide range in IP in the HETG spectra. To fit these, we included three zones ({\sc ATABLES}),
which we refer to as HIGHION, MEDION, and LOWION. MEDION produces most of the emission-lines and, therefore, we were able to constrain its 
parameters via the {\sc XSPEC} fitting. However, the fewer emission-lines from the other components, coupled with the weakness of the line fluxes compared to even the heavily absorbed
power-law, made it impossible to constrain their parameters. Hence, once we had obtained a good fit for MEDION,
we compared the line fluxes predicted by the Cloudy models to our measured fluxes to estimate $U$ and $N_{\rm H}$ for HIGHION and LOWION.
Then we added those components to optimise the fit to the emission lines, with the constraint that we could not over-predict features.
The final fitted parameters are listed in Table \ref{tab:Fits} and the individual model components are shown in Figure \ref{fig:ModComp}.  
Unless otherwise  stated, parameters in Table \ref{tab:Fits} are quoted in the rest-frame of the source and
errors are at the 90\% confidence level for one interesting parameter ($\Delta \chi^2 = 2.706$). For the final
model, the reduced $\chi^{2} = 1.028$, however, as noted above, the statistics were dominated by
the continuum fitting, rather than the emission lines or RRCs.

The values of $U$ and $N_{\rm H}$ for HIGHION and MEDION are similar to the two highest ionisation components used to fit the {\it XMM}/RGS spectra \citep{armentrout07a}.
LOWION has the same $U$ as the lowest ionisation component in \citet{armentrout07a}, but a significantly larger $N_{\rm H}$. 
At log$U = -0.5$, the strongest emission lines, such at the O~VII triplet, are formed near the ionised face of the slab, hence the fit to those
lines is not sensitive to the column density. However, there is some emission
from  inner shell Si~X - Si~V, in the 6.85 -- 6.92 \AA~ range (see Figure \ref{fig:MEG59}). These arise in LOWION, but would not have been
detected in the RGS spectra, which is likely why the value of $N_{\rm H}$ in \citep{armentrout07a} was so much lower. In any case, given the quality of these
data and the limited contribution from LOWION, its overall physical properties cannot be tightly constrained. 

\subsection{Initial Model Results}\label{sec:InitRes}

In Figures \ref{fig:MEG59} and \ref{fig:MEG917}, we compare the complete model to the HEG and MEG spectra, with the emission features
listed in Table \ref{tab:Lines} identified. In general, there is good agreement between the model and the data, in the 
sense that all of the strong emission lines and RRCs are present in the model. The only major
discrepancy is in the region near the Si~XIII triplet, at 6.6 - 6.8 \AA~in the observed frame.
In particular, there appears to be un-modelled emission at $\sim 6.7$\AA. In their analysis of the HETG spectra of the Seyfert 2 galaxy 
NGC~1068, \citet{kallman14a} identified this as Mg~XII, but the Mg~XII~$\beta$ line is relatively weak in their data, as well
as ours (see Table \ref{tab:Lines}), and this is at too low energy to be the Mg~XII RRC. Thefore, the source of the emission is unclear.

Although, overall the fit is reasonable, the model
unpredicts many of the emission lines. This can be rectified by increasing the nomalization of the {\sc ATABLES}, 
but to do so forces the model continuum to lie above the observed continuum. The limitation of the modelling is likely
due to unmodeled structure in the continuum.
For example, as shown in Figure \ref{fig:MEG59}, in the range from 7 - 8.5 \AA, the model flux lies above the data, and the 
latter shows structure consistent with unmodeled absorption. However, there is no obvious need for additional
emission components, or a broader range in $U$ and $N_{\rm H}$ for the {\sc ATABLES}. 

\subsubsection{Distance and Density Constraints}\label{sec:DistDense}

The {\sc ATABLE} grid was generated assuming a single density, $n_{\rm H} = 10^{5}$ cm$^{-3}$. We can achieve better
constraints on the model parameters by requiring $\Delta r/r < 1$ for each component, where $\Delta r$ and $r$ are 
the physical depth and distance from the ionzing source, respectively. This constraint can be reformatted in terms of $U$ and $N_{\rm H}$ 
such that:
  
\begin{equation}
r \leq Q/(4\pi c N_{\rm H} U)
\end{equation}
and
\begin{equation}
n_{\rm H} \geq (4\pi c N_{\rm H}^{2} U)/Q.
\end{equation}
Using the values of $U$ and $N_{\rm H}$ derived from the {\sc XSPEC} fitting, we computed values for $r$ and $n_{\rm H}$ for each component
which are listed in Table \ref{tab:Inputs}. We then generated new Cloudy models for each component. In Table \ref{tab:Inputs} we list the predicted average electron temperatures, $T_{e}$,
and Force Multiplers (FM), the ratio of the total radiation cross-section to the Thomson cross-section, at the ionised face and final zones,
for each model component\footnote{The ``ionised face'' refers to the initial Cloudy model zone, hence the part of a slab closest to the ionising source.
The ``final zone'' is the last zone computed by Cloudy in the determination of the radiative transfer through the gas.}.

The emitting area ($A$) of each component is computed from
the ratio of the emission-line and RRC luminosities, using the measured fluxes in Table \ref{tab:Lines} and assuming
a distance of 15 Mpc, to the Cloudy-predicted emission-line and RRC fluxes. Specifically,
$A =$ Luminosity/Flux. Additionally, we require the values of $A$ to be proportional to the relative normalisations of each component. Since 
the normalisations scale to the emission-line fluxes, we rescaled them based on the re-computed values of $n_{\rm H}$. And, since the values of $N_{\rm H}$ are fixed,
the fluxes and the normalisations scale as $n_{\rm H}$. Keeping the normalisation of HIGHION fixed, those of MEDION and LOWION were
increased by factors of 20.1 and 11.5, respectively.
The covering fraction of each component is calculated as $C_{f} = A/(4\pi r^{2})$.  
\citet{das05a} calculated a half-opening angle of the optical emission-line bicone in NGC~4151 of 33$^{\rm o}$. If the X-ray emission-line gas is confined to the bicone, $C_{f} \leq 0.3$.  Our predicted
$C_{f}$ are consistent with this scenario.   
The final values of $A$ and $C_{f}$ for each components are listed in Table \ref{tab:Inputs}. 

\subsubsection{Comparison to Emission Line Measurements}\label{sec:LineComp}

The model-predicted fluxes and luminosities of the emission-lines and RRCs are compared with the measured
values in Table \ref{tab:FinalFit}. The last column of the table is the data/model ratio. Most of the features are well-fit, with 0.56 $<$ data/model
$< 1.74$, with
the exception of the Si~XIII triplet, as noted above. The contributions of the individual components
are clearly distinguished, with emission from ions with IP, $>$ 1.2 keV, e.g. Si~XIV, Si~XIII, and Mg~XII, coming from HIGHION,
those with the lowest IP, e.g. O~VII (138.08 eV), coming from LOWION, and MEDION contributing
to all ionisation states. 

As noted in Section \ref{sec:Lines}, we were unable to obtain accurate constraints on the widths of the RRCs. However, we were able to determine if the predicted electron temperatures ($T_{e}$) were consistent with the data. To do so,  we refit the O~VII and O~VIII RRCs with a fixed ratio 
of the contributions from each model component (see Table \ref{tab:FinalFit}) and fixed $kT_{e}$, where $k$ is Boltzmann's constant. For O~VII, assuming 90\% at $T_{e}=1.7 \times 10^{5}$K, from MEDION,
and 10\% at $T_{e} =1.9 \times 10^{4}$K, from LOWION, this yielded 
a normalization 8.41e-05 with 90\% confidence range from 5.99e-05 - 1.11e-04.
For O~VIII, assuming 67\% at $T_{e}=2.94 \times 10^{6}$K, from HIGHION, and 33\% from MEDION, this yields a  normalization of 1.33e-04  with 90\% confidence range 1.02e-04 - 1.65e-04. Hence the model predictions are consistent with the data, but
the values are not well constrained. However, there is no evidence that the values are significantly higher, especially
for O~VII, hence there is no need for a collisionally-ionised component in the spectral fitting.

With the large column densities required by the fit to the HETG spectra, the models also predict
significant UV line emission, the strongest of lines being O~VI $\lambda\lambda$1032, 1038. 
The combined O~VI luminosity from the models is 
$L_{OVI} = 1.9 \times 10^{40}$ erg s$^{-1}$, $\sim$ 85\% of which is from LOWION. Based on {\it FUSE} spectra, the O~VI flux from the NLR
is $2 \times 10^{-12}$ erg cm$^{-2}$ s$^{-1}$, or $L_{OVI} =  5.4 \times 10^{40}$ erg s$^{-1}$ \citep{armentrout07a}.
Therefore, while the X-ray gas is predicted to have a strong UV signature, the models are consistent with the UV emission-line 
constraints. 
 
\subsubsection{Masses and Mass Outflow Rates}\label{sec:MMdot1}
 
The mass of emission-line gas for each component is $M = \mu m_{p} A N_{\rm H}$,
where, $m_{p}$ is the mass of a proton, and $\mu =1.4$, which accounts for the mass of helium and the
heavy elements, at roughly solar abundances. This may also be written as $M = 4 \pi r^{2} C_{f} \mu m_{p} n_{\rm H}$ $\Delta r$,
where $n_{\rm H} \Delta r  = N_{\rm H}$.  For the three model components, we obtain the following:
HIGHION, 118 M$_\odot$; MEDION, 2.59 $\times 10^{3}$ M$_\odot$; and LOWION, 4.17 $\times 10^{3}$ M$_\odot$.  
The mass outflow rate is $\dot{M}_{out} = 4\pi r N_{\rm H}\mu m_{p} C_{f} v_{r}$. Assuming the
average value of $v_{r}$ given in Section \ref{sec:LineVels}, we obtain the following: HIGHION, 0.03 M$_\odot$ yr$^{-1}$;
MEDION, 0.11 M$_\odot$ yr$^{-1}$; and, LOWION, 0.09 M$_\odot$ yr$^{-1}$. Although not strictly lower limits,
these values were computed assuming the components were at distances computed by requiring $\Delta r/r < 1$, and we did not 
use a deprojected $v_{r}$. The summed mass outflow rate is 0.23 M$_\odot$ yr$^{-1}$, which is close to the optical/UV\footnote{The term Optical/UV refers to the
values derived from the analysis of STIS spectra covering its full UV to near IR bandpass.} rate
from within 12 pc of the nucleus, $\approx$ 0.3 M$_\odot$ yr$^{-1}$ computed by \citet{crenshaw15a}.
However, given that several of the emission lines extend more than 100 pc from the nucleus (see Figure \ref{fig:Zeroth}) and the
deprojected velocities are likely higher than $v_{r}$ \citep{crenshaw15a}, the values of $M$ and $\dot{M}_{out}$ can be significantly
greater. We address this in the next section.  

\section{Analysis of Extended Emission}\label{sec:ExtEmm}

\subsection{Zeroth Order Line Profiles}\label{sec:Zeroth}

The soft X-ray emission is resolved in NGC~4151 \citep{ogle00a} and extends beyond 1 kpc \citep{wang11a}. As discussed in Section 2.1, the zeroth-order image shows H- and He-like O and Ne lines extending to $\geq$ 100 pcs from the nucleus. For the extended gas to be in a sufficiently
high state of ionisation to produce these lines, the densities must decrease with distance, as suggested by \citet{wang11b}. Line emissivities are proportional to $n_{\rm H}^{2}$ \citep{osterbrock06a}. Therefore, even though the line fluxes decrease with distance, there may be a significant amount of mass associated with the extended emission.   

While in NGC 1068, there are enough counts in the $\pm$ 1$^{\rm st}$-order HETG spectra to obtain accurate emission line fluxes outside the inner nucleus (see \citealt{kallman14a}), this is not the case for NGC~4151. Hence, we used the zeroth-order spectral images to measure the extended emission. We measured the fluxes in regions of 0\arcsecpoint5 in length and 3\arcsecpoint0 in width, centered at the nucleus, along a position angle of 140\degree. Our extraction    
window is shown, superimposed on an archival {\it HST}/WFPC2 [O~III] image, in the left hand panel of Figure \ref{fig:XBOX}. The right hand panel of Figure \ref{fig:XBOX} shows the [O~III] emission profile over the same window.  In Figure \ref{fig:NEIXNEX}, we show the extracted Ne~IX rif and Ne~X $\alpha$ profiles, compared with the 2.05-2.25 keV continuum, which was chosen becasue it is free of strong line emission. The emission line profiles are clearly broader than the continuum and show structure that is similar to the [O~III] profile. 

We determined the fraction of emission in each 0\arcsecpoint5 bin by dividing counts in the bin by
the total counts for each line. These were converted to line luminosities as a function of distance
by multiplying by the total luminosity of the lines, listed in Table \ref{tab:FinalFit}. In our analysis, we only considered bins for which
the signal-to-noise ratios were $>$ 2. 

\subsection{Estimating the Mass of the Extended Gas}\label{sec:MMdot2}

In order to fit the spatial resolved emission-line profiles, we generated Cloudy models
for distances corresponding to the center-points of the 0\arcsecpoint5 bins. The 
distances from the ionising source for each model were determined assuming an inclination angle of 45\degree, which is the
inclination of the bicone axis with respect to the plane of the sky 
\citep{das05a}. Because we
are mapping Ne~IX rif and Ne~X $\alpha$, we only ran these models
for the HIGHION and MEDION\footnote{We chose to map the Ne~X and Ne~IX emission, rather than O~VIII and O~VII, because the
latter are at lower energies, where the decrease in ACIS sensitivity is more severe, and are strongly dependent on the predictions from the LOWION,
which is not well-constrained.}. In doing so, we assumed that density decreases as
$n_{\rm H} \propto r^{-1.65}$, which is the same as for the optical/UV emission-line gas used in \citet{crenshaw15a}.
\citet{crenshaw15a} based their density law on our analysis of {\it HST}/STIS spectra \citep{kraemer00b}, in we used density diagnostics, such as the [S~II] 6716\AA/6731\AA~ratio,
and constraints from photoionisation models.  
The density laws for each
component were normalised using the densities obtained for the global models (Table \ref{tab:Inputs}).
Although there is no direct evidence for such a density law for X-ray gas, there are
two reasons why it is plausible. Firstly, the fact that the gas is being mapped out by Ne~IX and Ne~X
suggests that the ionisation state of the gas is not dropping precipitously, which implies
that the density could fall as $r^{-2}$, as proposed by \citet{wang11b}. Secondly,
as we will discuss in the next section, the X-ray gas may arise from expansion of the
optical emission-line gas, which could lead to the same density law for both.
We constrained $N_{\rm H}$ by requiring $\Delta r/r <1$, as before, although for radial distances $>$ 25 pc, $r$ is the 
deprojected size of the extraction bin.
Hence, $N_{\rm H}$ decreases with $r$, 
as $n_{\rm H}$ is decreasing. The Cloudy model parameters are listed in Table \ref{tab:SpatRes}.

In order
to calculate the masses in each bin, we used the line luminosities and the Cloudy-predicted fluxes for Ne~IX rif and Ne~X$\alpha$
to solve for the emitting areas of each component, $A_{\rm MED}$ and $A_{\rm HIGH}$, using the following equations:

\begin{equation}
F(NeIX)_{MED} \times A_{\rm MED}~+~F(NeIX)_{HIGH} \times A_{\rm HIGH} = L_{\rm NeIX} \times frac9
\end{equation} 
and,
\begin{equation}
F(NeX)_{MED} \times A_{\rm MED}~+~F(NeX)_{HIGH} \times A_{\rm HIGH} = L_{\rm NeX} \times frac10
\end{equation} 
where $F(\rm NeIX)_{MED}$, $F(\rm NeIX)_{HIGH}$, 
 $F(\rm NeX)_{MED}$, and $F(\rm NeX)_{HIGH}$ are the Cloudy-predicted fluxes from each component,
$frac9$ and $frac10$ are the fractions of the emission-line fluxes in each bin, and $L_{\rm NeIX}$
and $L_{NeX}$ are the line luminosities (see Table \ref{tab:FinalFit}). 

In order to check that the set of models for the extended emission still provided a good fit to the
spectra, we used the values for $A_{\rm MED}$ and $A_{\rm HIGH}$ to compute the luminosities of other
emission lines. The predictions for the Mg~XI r and f and Si~XIII r lines remained well-within the measurement uncertainties.
The fit for Si~XIII f worsened (data/model $=$ 5.3), but the line was not well fit by the initial model
(see Section \ref{sec:LineComp}). However, the fit for Mg~XII $\alpha$ was now poor (data/model $=$ 2.3), and the new
models predicted only 15\% of the Si~XIV $\alpha$. Nevertheless, these results are consistent with the unresolved profiles
for these lines (see Figure \ref{fig:Zeroth}). In summary, while the models provide a good fit to the resolved emission lines,
there is still need for a component of high-ionisation gas close to the AGN.

Once we determined the $A$'s for each component, we calculated $M$ for each bin using the
model $N_{\rm H}$ (Table 5), as described in Section \ref{sec:MMdot1}. The final values are plotted in the left hand panel of Figure \ref{fig:MMdot}.
The low value for $M$ in the centre bin is due to the high density (see Table \ref{tab:SpatRes}). The reason for this is that the emissivity 
of the gas goes as $n_{\rm H}^{2}$, hence, less mass can produce the same emission.
The uncertainties in the X-ray masses are due to photon-counting statistics, while those in position are
due to the size of the extraction bins. The line emissivities are proportional, hence the values of $A$ are inversely proportional,
to $n_{\rm H}$ (see \citealt{osterbrock06a}. The total mass of the ionised gas is $5.4 (\pm 1.1) \times$ 10$^{5}$ M$_\odot$.

As shown in \citet{crenshaw15a} and \citet{revalski18b}, the optical outflows in NGC~4151 extend out to a distance of $\sim$ 150 parsecs. To compare the mass
distribution of the X-ray emission-line gas to that of the optical/UV gas within the outflow region, we summed the X-ray masses from the same distances to the NE and SW 
of the nucleus. We summed the masses from optical analysis to correspond to the bin sizes used for the X-ray extraction; the errors in the mass values are those
from \citet{crenshaw15a}, added in quadrature to account for the binning. The results
are shown in the right hand panel of Figure \ref{fig:MMdot}.  The totals are roughly similar. Note that the drop in optical/UV mass in most distant zone
is primarily an artifact, resulting from the truncation of the optical measurement at 132 pc. The total mass of X-ray gas within the outflow regions is
$3.9 (\pm 0.4) \times 10^{5}$ M$_\odot$, compared to $\approx$ $3 \times 10^{5}$ M$_\odot$ for the optical/UV component.   

\subsection{Mass Outflow Rates in the Extended Gas}\label{sec:MMdot3}

Using the derived masses for the X-ray gas, we calculated the values of $\dot{M}_{out}$ as a function of distance,
as described in Section \ref{sec:MMdot1}. However, since the physical depth, or $\Delta r$, of each model component is now limited by the size of the extraction bins,
rather than requiring $\Delta r/r < 1$, the equation for the mass outflow rate can be rewritten as
$\dot{M}_{out} =  \frac{M v_{r}}{\Delta r}$ \citep[e.g.,][]{revalski18a}.
In computing $\dot{M}_{out}$, we assumed that the X-ray gas exhibited the same kinematics as the [O~III]
gas. There are two main reasons as to why this is a plausible assumption.  First, as noted in Section \ref{sec:LineVels}, while the average velocity
centroid for the X-ray lines is higher than that of the collapsed [O~III] centroid, this is likely due to absorption of the soft X-ray emission NE of
the nucleus. Second,
as shown in the Figures \ref{fig:XBOX} and \ref{fig:NEIXNEX}, the profiles of the [O~III], Ne~IX, and Ne~X lines are similar when measured
over the same extraction bins.   

In Figure \ref{fig:MDOTCOMP}, we show  $\dot{M}_{out}$ as a function of de-projected radial distance, for both the X-ray and optical/UV gas.
The errors plotted for the X-ray $\dot{M}_{out}$'s, in addition to the contribution from the mass estimates, include those of the
[O~III] velocity measurements. The errors for the optical/UV points are from \citet{crenshaw15a}, again added in quadrature to account for
the binning. The X-ray $\dot{M}_{out}$'s are similar to those in the optical/UV gas, although the former does not drop off at the highest distance.
Note that the smaller size for the final bin in the optical/UV data also enters into the equation for $\dot{M}_{out}$, in a manner that
compensates for the smaller mass in that bin (see Section \ref{sec:MMdot1}) hence the drop-off in $\dot{M}_{out}$ is not an artifact in this case.
The maximum X-ray $\dot{M}_{out} = 1.80 \pm 0.57$ M$_\odot$ yr$^{-1}$, as compared to $ 2.31 \pm 0.67$ M$_\odot$ yr$^{-1}$ for the optical/UV gas.
Note the latter is somewhat smaller than the $\sim 3$ M$_\odot$ yr$^{-1}$ given in \citet{revalski18b}; this is due to the averaging of the velocity
within the extraction bin.  While the outflow rates are similar, the possibility of a greater extent for the X-ray outflows suggests that its
importance in mass outflow, hence AGN feedback, may have not been fully appreciated, as we discuss in Section \ref{sec:DISC}.

From these results, we determined a peak $L_{KE} = 2.92 (\pm 1.22) \times 10^{41}$ erg s$^{-1}$, at a distance of 150 pc.
This is similar to the value derived by \citet{wang11b} and confirms that the power of the X-ray outflow is comparable to that 
of the optical/UV gas \citep{crenshaw15a}.
Adding in the optical/UV contribution, $L_{KE}$ is $\sim$ 0.5\% of L$_{bol}$, which is
at the limit for efficient feedback of 0.5\%-5\% \citep{hopkins10a}, although these limits may not be relevant for such a sub-Eddington AGN
\citep{king15a}.

\section{Discussion}\label{sec:DISC}

As shown in Section \ref{sec:ExtEmm}, the extended X-ray emission-line gas in NGC~4151 may possess similar
mass, mass distributions, and mass outflow rates as the optical/UV gas. 
However, the low value of $L_{edd}/L_{bol}$ in NGC~4151 calls into question the physical process by which the gas is accelerated.
For example, the possibility of radiative acceleration of the gas can be determined by comparing the predicted $FM$s (see Table \ref{tab:Inputs}) 
to $L_{edd}$/$L_{bol}$. Specifically, radiative acceleration is proportional to $FM \times L_{bol}$ \citep[e.g.,][]{das07a}.
For $FM = 1$, the only source of opacity is electron scattering, via the Thomson cross-section, and radiative acceleration can only occur
if the source is radiating at $L_{edd}$. At lower ionisation, the opacity and, hence, $FM$ increases. Therefore, radiative 
acceleration can occur in sources with $L_{bol} = L_{edd}/FM$.  Using our estimate for $L_{bol}$, radiative acceleration requires $FM \geq 25$.
Therefore, radiative acceleration is not possible for HIGHION, while it would marginally possible for MEDION if $L_{bol}$ is closer to the value
calculated by \citet{wang11b}. LOWION dynamics are consistent with radiative acceleration, but the large difference in FM across
the zone suggests that LOWION cannot be dynamically stable (and its physical characteristics are not well-constrained by our models). 

An alternative means of acceleration for highly ionised gas, in sub-Eddington AGN, is via magneto hydrodynamic (MHD) processes. In fact, MHD has been suggested as the acceleration
mechanism for the most highly
ionised absorbers in NGC~4151 \citep{kraemer05a, kraemer18a}. Effective MHD requires strong magnetic fields, such as those associated with
the accretion disk \citep{blandford82a}, which infers that the outflows originate close to the AGN. However, \citet{crenshaw15a} argued that
there was significantly more optical/UV gas in the NLR of NGC~4151 than could be supplied by outflows from the inner nucleus. This is
almost certainly the case for the X-ray gas, given the similar
masses and mass outflow rates to the optical/UV component. Instead, they suggested
that most of the gas has been accelerated in situ, and originated in dust spirals that had crossed the illumination cone. This scenario has been
explored in more detail by \citet{fischer17a}, who demonstrated that outflows in the NLR of the Seyfert 2 Mrk 573 could be explained by in situ 
accleration of material in the host galaxy's disk. As such, it is unlikey that the X-ray gas is part of an MHD flow. 

If the bulk of the extended X-ray gas originates in the host galaxy's disk, and was accelerated in situ, the $FM$'s listed in Table \ref{tab:SpatRes} 
are more relevant for evaluating the efficacy of radiative acceleration. However, at these larger radial distances, the enclosed stellar mass 
becomes a dominant source of gravitational deceleration, as shown in \citet{das07a}. Based on their results, a $FM \sim 50$ is far too low
to launch an outflow in NGC~1068, which is more luminous than NGC~4151
(see \citealt{kraemer15a}). Therefore, radiative acceleration would be unlikely
even if the outflows were launched at a time when NGC~4151 was radiating at closer to its Eddington limit.

Embedded dust may play an important role in AGN-driven winds \citep[e.g.,][]{ishibashi18a}. Specifically, dust grains may provide
sufficient opacity for radiative acceleration of highly ionised winds in sub-Eddington AGN. To test this, we generated models for the MEDION component which included
a Galactic interstellar medium dust fraction, using the ``ISM'' option \citep{ferland17a}. Under these conditions, the
Cloudy models predict $FM \sim 300$. Using the expression for radial velocity as a function of distance given in \citet{das07a} and
their enclosed mass profile\footnote{\citet{das07a} used the mass profile for NGC~1068; since NGC~4151 has been classified
as an earlier-type spiral galaxy than NGC~1068 \citealt{devaucouleurs91a}, it is likely that the enclosed mass at the same radial distance is greater.}, 
we determined that dusty winds cannot be effectively launched at distances $\gtrsim$ 10 pc. Hence,
while dust may play an important role in the acceleration of highly ionised close to the AGN, the associated increase in $FM$ is not sufficient 
to account for the generation of winds further into the host galaxy.    

Another possibility is that the X-ray gas was in a lower state of ionization, by virtue of being higher
density, when initially accelerated. Indeed, this has been suggested for the radiative acceleration of Ultra Fast Outflows in AGN \citep{hagino15a}.
One scenario is that the X-ray gas arises as the optical/UV gas expands. Photoionised gas in which [O~III] is a dominant coolant
would have $T_{e} \sim 15000$K \citep{osterbrock06a}, which results in a sound speed $\sim 2 \times 10^{6}$ cm s$^{-1}$. Unless somehow
confined, the gas would become highly rarefied close to its point of origin. Based on the results from Mrk 573 \citep{fischer17a}, it does appear
that knots of [O~III] emission can travel 10s of pcs, which suggests some confinement, but the lack of high velocity gas beyond
several 100 pcs from the AGN is evidence that these knots eventually expand to the point that they are too highly ionised to be detected in [O~III].
Therefore, it is plausible that the X-ray outflow arises from optical/UV gas that is accelerated in situ.

\section{Conclusions and Summary}

We have analysed {\it Chandra}/HETG spectra of NGC~4151. The data were taken with an orientation to ensure that the
axis of the NLR bicone was perpendicular to the dispersion direction, which enables us to map the
extended emission-line structure. Our main conclusions are as follows.

1. In the zeroth order spectral image, the continuum and lines with energies $>$ 1.3 keV are compact
and centred on the nucleus. The Si~K$\alpha$ and Fe~K$\alpha$ lines are also
strongly centrally peaked, consistent with an origin in the putative torus. On the other hand,  
the Ne~X, Ne~IX, O~VIII, and O~VII lines  are all clearly extended along a NE-SW direction, out to an undeprojected distance of $>$ 200 pc.
This is in agreement with previous {\it Chandra} studies \citep[e.g.][]{ogle00a, wang11a} and indicates that much
of the soft X-ray emission arises in the NLR.

2. The emission lines have an average radial velocity $v_{r} = -230^{-90}_{-370}$ km s$^{-1}$, with no clear
trend with ionisation potential. This is higher than the value obtained from the collapsed STIS spectra \citep{crenshaw10a},
which we attribute to absorption of red-shifted soft X-ray emission, NE of the nucleus, either by gas in the disk of the host galaxy or
material associated with X-ray absorbers. Taking that into account, it is plausible that optical and X-ray outflows have similar kinematics.

3. We were able to fit the 1st order HETG spectra with a model consisting of a reflection component,
and absorbed power-law, with a small fraction of unabsorbed continuum, and three emission components. The absorption 
and emission components, or {\sc MTABLES} and {\sc ATABLES}, were generated with the photoionisation code Cloudy
\citep{ferland13a, ferland17a}.
The models for the {\sc ATABLES} were parameterised with log$N_{\rm H}$/log$U$ of 22.5/1.0, 22.5/0.19, and 23.0,-0.50, and
named HIGHION, MEDION, and LOWION, respectively. HIGHION and MEDION are similar to the two highest ionisation components
used to fit the XMM/RGS spectra \citep{armentrout07a}. On the other hand, LOWION, while required by the spectral fitting, is
not well-constrained, primarily due to the lack of sensitivity of {\it Chandra} at the lowest energies.

4. As shown in Table \ref{tab:FinalFit}, the emission line fluxes predicted by the Cloudy models provide
a good fit to the observed emission lines. Hence, we used the models as the basis for analysing the extended emission.
To model the extent, we used the zeroth order emission-line profiles for Ne~IX and Ne~X (see Figure \ref{fig:NEIXNEX}) and assumed that the
X-ray gas followed the same density law as the optical/UV gas \citep{crenshaw15a}, or $n_{\rm H} \propto r^{-1.65}$. 
Masses were determined for each emission component within 50 pc bins, to $r > 200$ pc. The summed masses are shown in Figure \ref{fig:MMdot}.
We derived a total mass of $5.4 (\pm 1.1) \times$ 10$^{5}$ M$_\odot$. 

5. By assuming the same velocity profile derived from the STIS analysis, we were able to use the models for the extended
emission to compute $\dot{M}_{out}$ as a function of $r$. We obtain a maximum X-ray $\dot{M}_{out} = 1.80 \pm 0.57$ M$_\odot$ yr$^{-1}$, 
as compared to $ 2.31 \pm 0.67$ M$_\odot$ yr$^{-1}$ for the optical/UV gas. Our value is close to that derived by \citet{wang11b} in
their {\it Chandra} imaging analysis. Note, we did not include the contribution from LOWION, hence the true X-ray mass outflow rate may
be somewhat higher.

6. Although the kinematics of the X-ray gas appears to resemble the optical/UV gas, its relatively high ionisation state
calls into question the efficiency of radiative acceleration. Although the presence of dust mitigates this, the corresponding
increase in cross-section is not sufficient to overcome the gravity of the enclosed stellar mass, presuming, as we argue, most of the gas is launched
in situ, in the plane of the host galaxy. One possible explanation is that the optical/UV gas is not fully confined. Hence, the X-ray 
component arises from thermal expansion of lower-ionization, more efficiently accelerated gas.

In summary, based on our results, the X-ray emission line gas has a similar mass and peak mass outflow rate
as the optical/UV gas. Therefore, it is an important component in AGN-driven winds. Furthermore, it does not
appear to exhibit a drop in $\dot{M}_{out}$ for $r > 100$ pc. That may indicate that the X-ray component has a 
greater effect on the host galaxy  than the optical/UV gas and that X-ray winds might be a more efficient mechanism for AGN feedback. It would be interesting to see if there is a similar, or greater, role for X-ray winds in higher luminosity AGN.

\section*{Acknowledgements}

Support for this work was provided by the National
Aeronautics and Space Administration through Chandra Award Number G04-15106X
issued by the Chandra X-ray Observatory Center, which is operated by the
Smithsonian Astrophysical Observatory for and on behalf of the National
Aeronautics Space Administration under contract NAS8-03060, and from
program GO-13508, support for which was provided by NASA through a grant from
the Space Telescope Science Institute, which is operated by the Association of
Universities for Research in Astronomy, Inc., under NASA contract NAS5-26555.
This research has made use of data and/or software provided by the High
Energy Astrophysics Science Archive Research Center (HEASARC), which is a
service of the Astrophysics Science Division at NASA/GSFC and the
High Energy Astrophysics Division of the Smithsonian Astrophysical Observatory.We are  grateful to the {\it Chandra} operations teams  for performing the observations  and providing 
software and calibration for the data analysis.
We thank Keith Arnaud, for his continuuing maintenance and development of
XSPEC, and Gary Ferland and associates, for the maintenance and development of
Cloudy.

\begin{figure*}
\includegraphics[trim= 0 40 0 0,clip, width=18cm, height=15cm, angle=0]{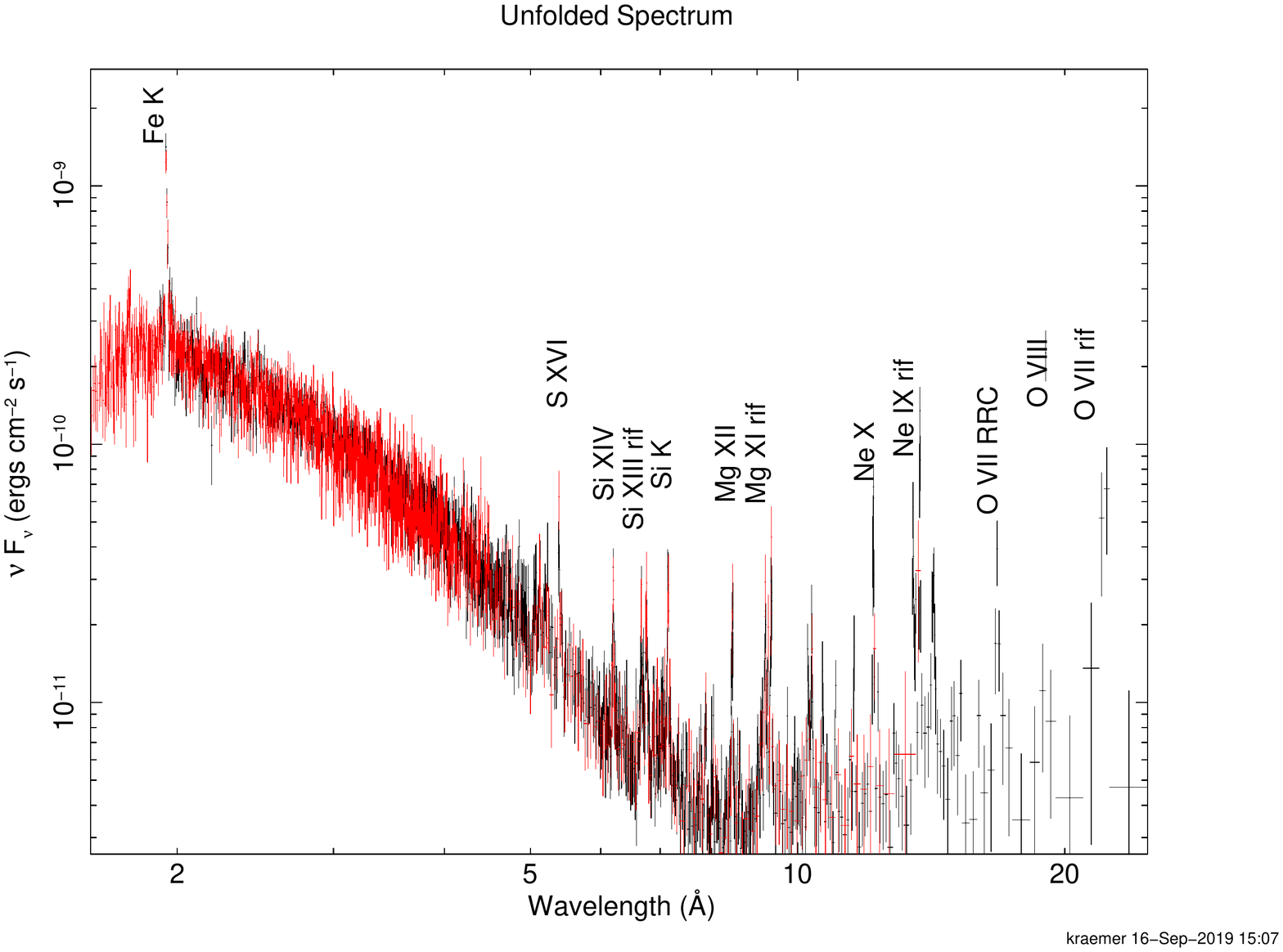}
\caption{Full, $\pm$ 1$^{\rm st}$ order, HETG spectrum from 1.59 \AA - 24.8 \AA, from the summed OBSIDs (see \ref{sec:Spectra}).
The red and black points are from the HEG
and MEG, respectively. The data were binned to a minimum of 10 counts (per bin). The turn-over   
towards lower energies (greater wavelengths) is due to intrinsic absorption (see \citealt{couto16a}). Emission at wavelengths
$>$ 6 \AA~is emission-line dominated.} 
\label{fig:FullSpec}
\end{figure*}

\begin{figure*}
\includegraphics[width=18cm, height=15cm, angle=0]{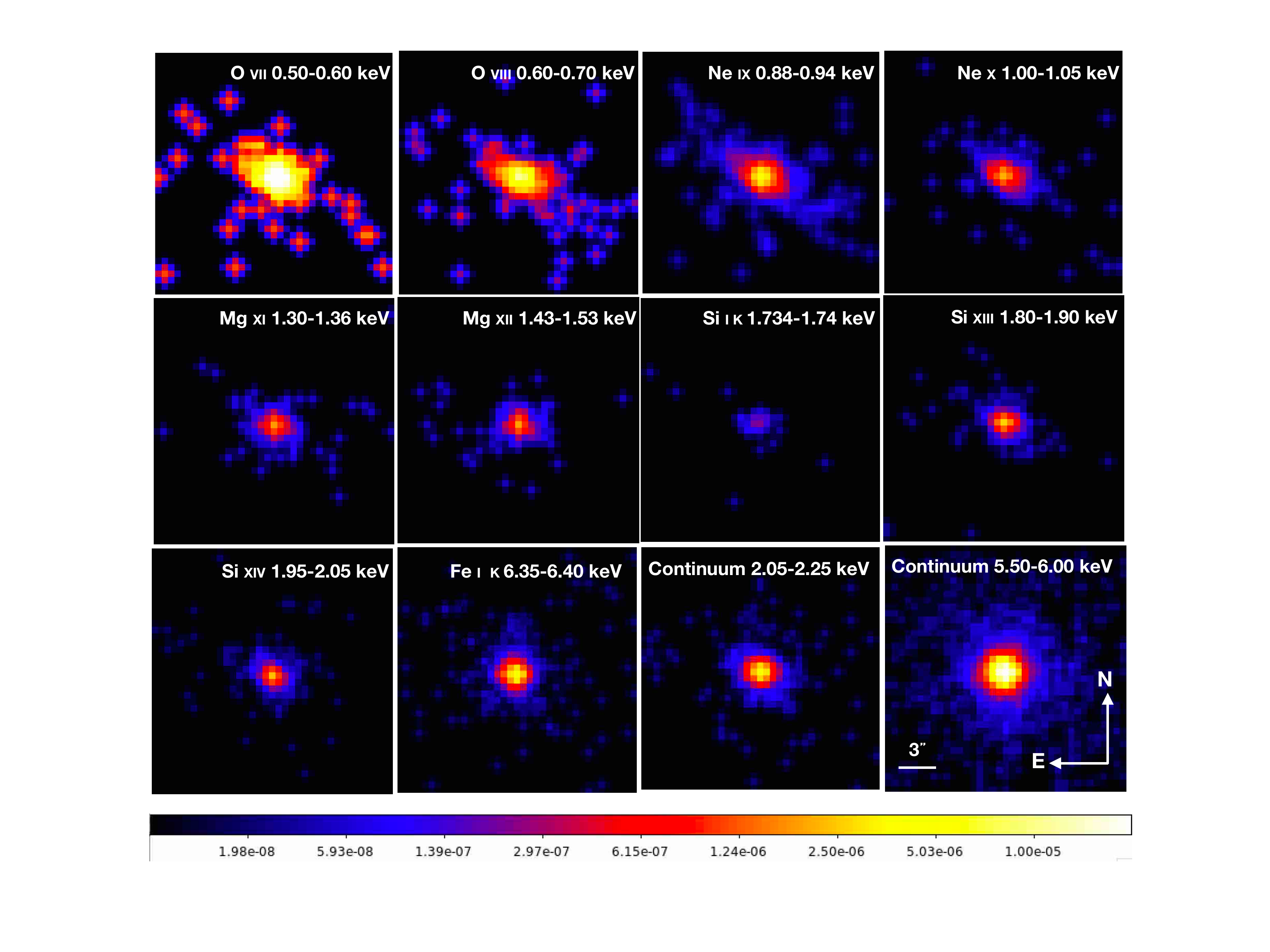}
\caption{Emission profiles for hydrogen and helium-like O, Ne, Mg, and Si lines, and Si~K$\alpha$ and Fe~K$\alpha$ lines, from the zeroth-order HETG spectra.
Line-free regions of continuum in the ranges 2.05-2.25 keV and 5.50-6.00 keV are also shown. The images have been smoothed with a Gaussian function with kernel
radius of 1$^{''}$, and are equally scaled for comparison in the colorbar in terms of total number of counts. The spatial scale and orientation are shown in the lower right panel.
} 
\label{fig:Zeroth}
\end{figure*}

\

\begin{figure*}
\centering
\includegraphics[width=8.5cm, angle=180]{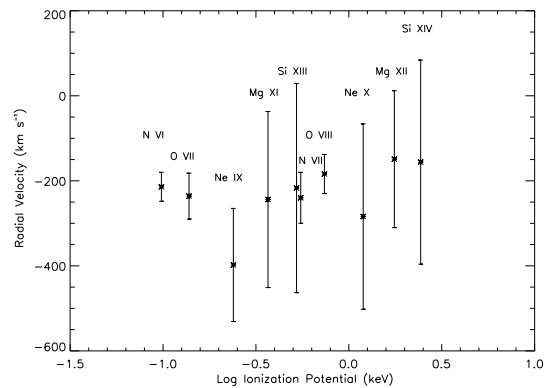}
\caption{Radial velocities plotted versus ionisation potential (IP). Velocities of H-like ions were measured from their Ly-$\alpha$ transitions, while those for the He-like ions were measured from the triplet forbidden lines. Velocities from N~VI, N~VII, O~VII, O~VIII, and Ne~IX were taken from the {\it XMM}/RGS analysis
\citep{armentrout07a}. IPs not listed in Table \ref{tab:Lines} are: N~VI, IP $=$ 0.098 keV; N~VII, IP $=$ 0.55 keV \citep{cox00a}.} 
\label{fig:Velocities}
\end{figure*}

\begin{figure*}
\centering
\begin{minipage}[b]{.40\textwidth}
\includegraphics[trim= 0 40 0 0,clip, width=8cm]{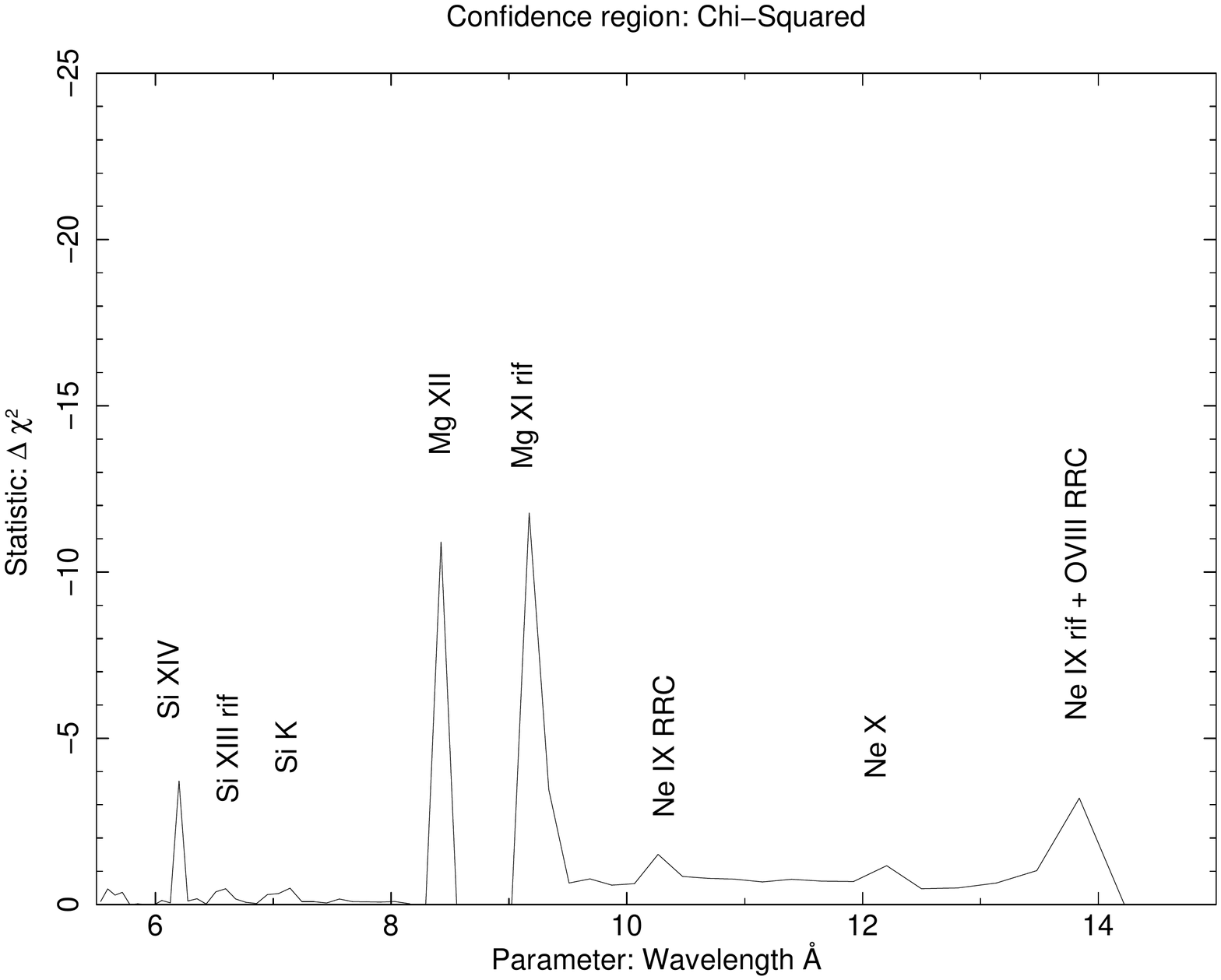}
\end{minipage}\qquad
\begin{minipage}[b]{.40\textwidth}
\includegraphics[trim = 0 40 0 0,clip,width=8cm]{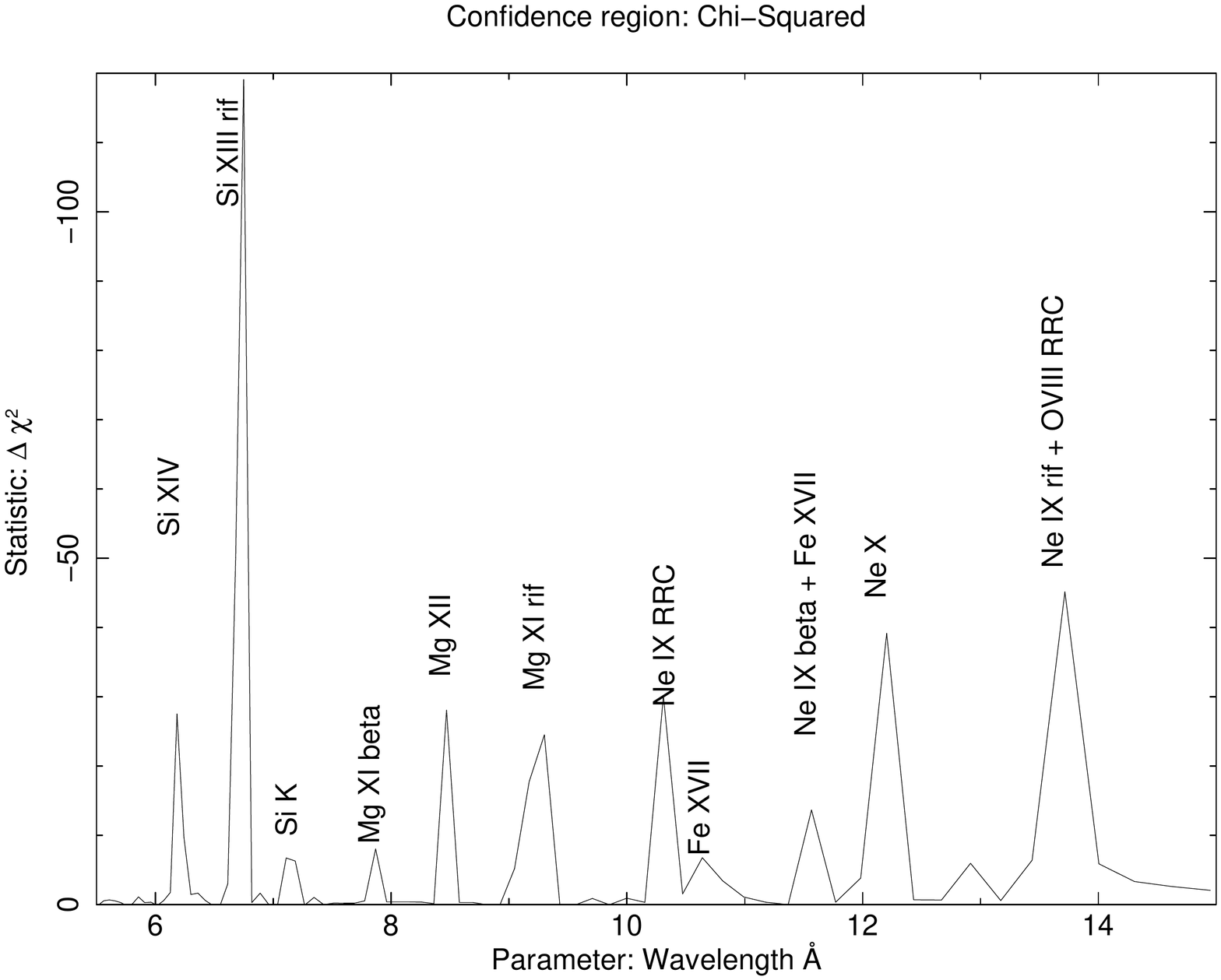}
\end{minipage}
\caption{Sliding Gaussians tests, over the range 5.5\AA-15\AA, for the HEG (left hand panel) and MEG (right hand panel)
spectra. The test was performed on the spectra with the absorbed power-law and reflection
components included (see Section \ref{sec:Inputs}). The detected emission features are labeled.}\label{fig:Slide}
\end{figure*}

\begin{figure*}
\centering
\begin{minipage}[b]{.40\textwidth}
\includegraphics[trim= 0 40 0 0,clip, width=8cm]{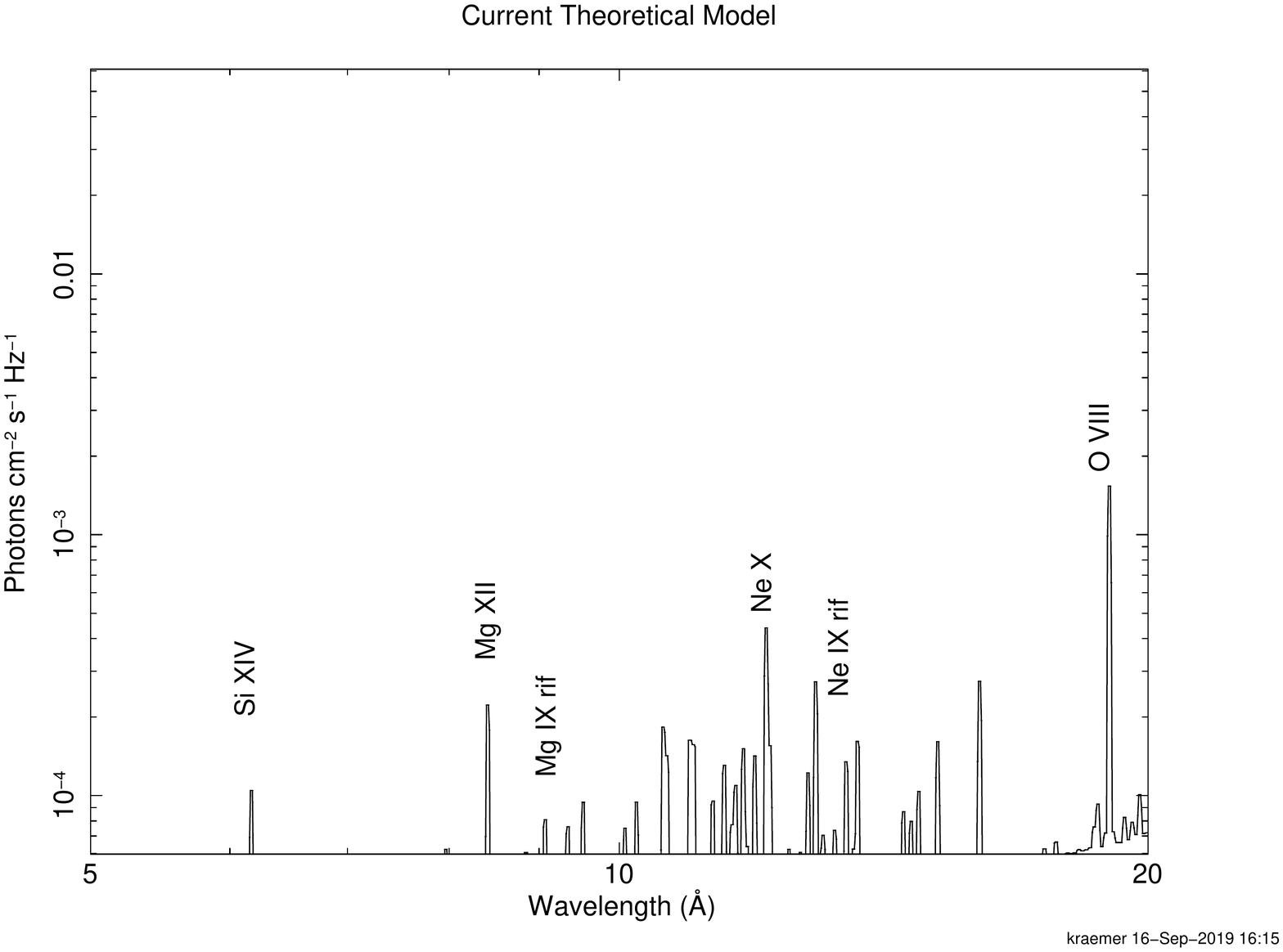}
\end{minipage}\qquad
\begin{minipage}[b]{.40\textwidth}
\includegraphics[trim = 0 40 0 0,clip,width=8cm]{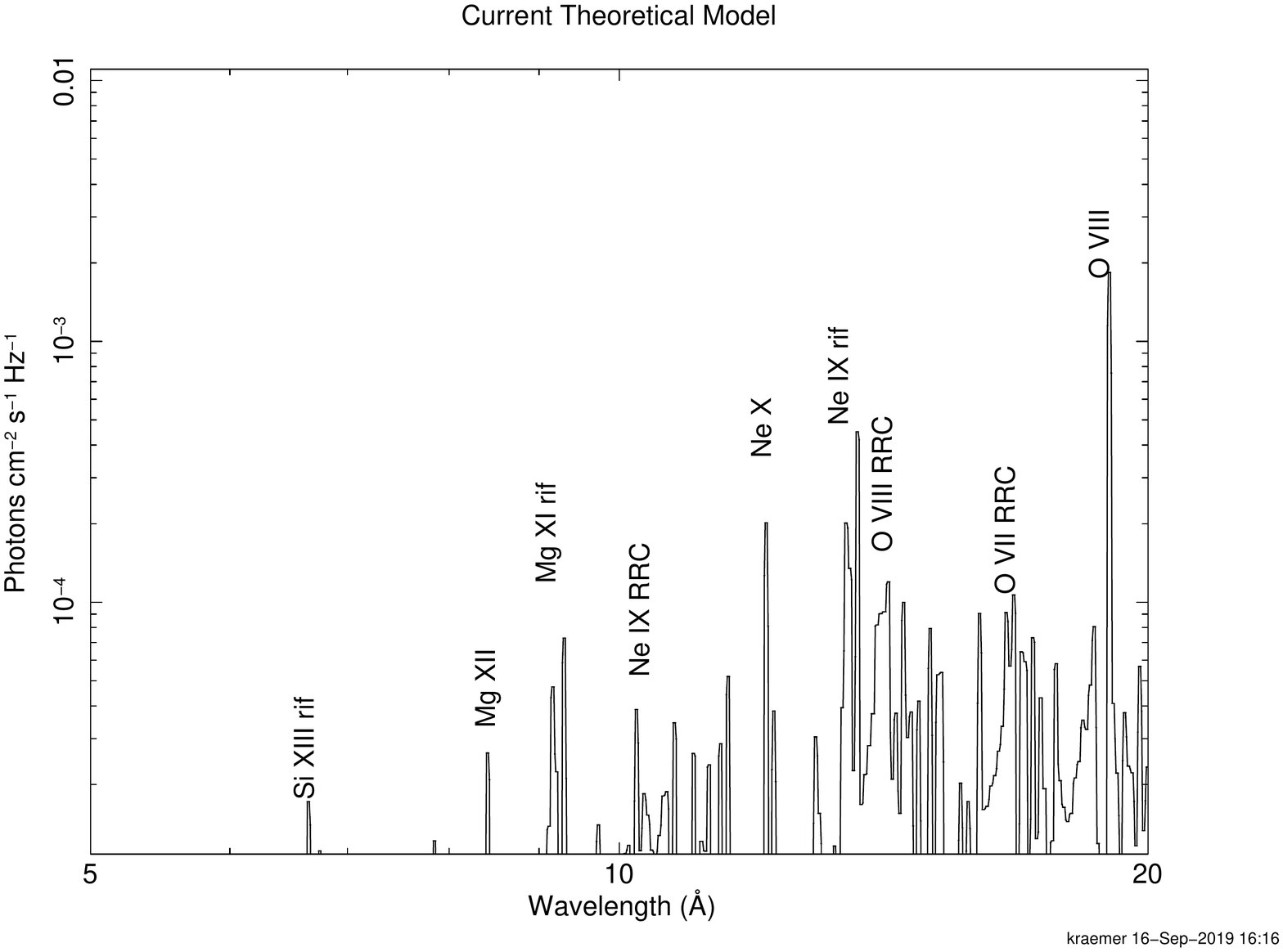}
\end{minipage}\qquad
\begin{minipage}[b]{.40\textwidth}
\includegraphics[trim= 0 40 0 0,clip, width=8cm]{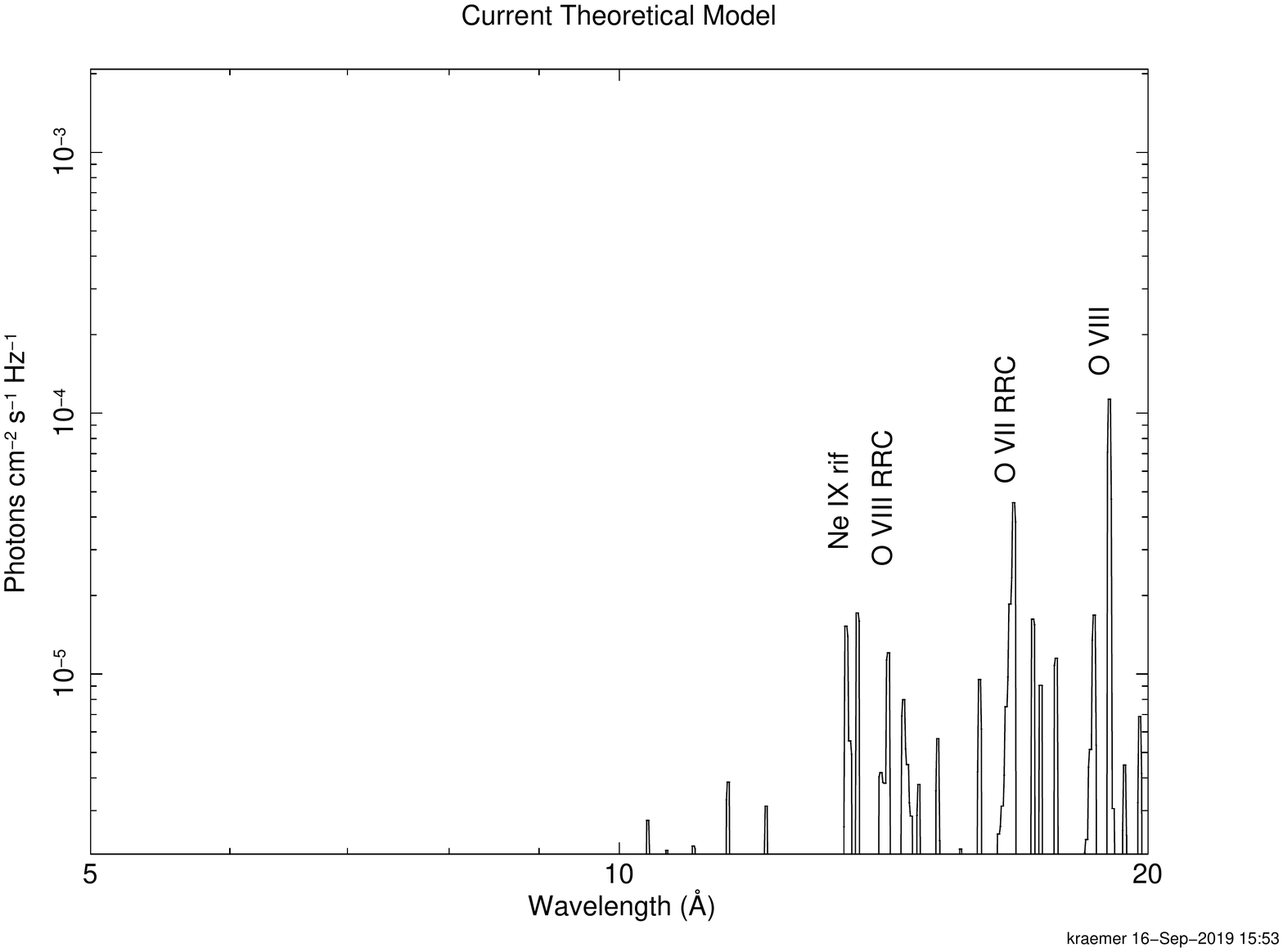}
\end{minipage}
\caption{Predicted emission-line spectra in the range 5\AA~ - 20\AA, for the three
model components described in the text: HIGHION (left hand panel), MIDION (right hand panel), and
LOWION (lower panel). The plots were generated from {\sc ATABLES} created with the
Cloudy model grid. Emission features used in the model fitting are labeled.}\label{fig:ModComp}
\end{figure*}

\begin{figure*}
\centering
\begin{minipage}[b]{.4\textwidth}
\includegraphics[trim= 20 40 0 0,clip, width=8.5cm]{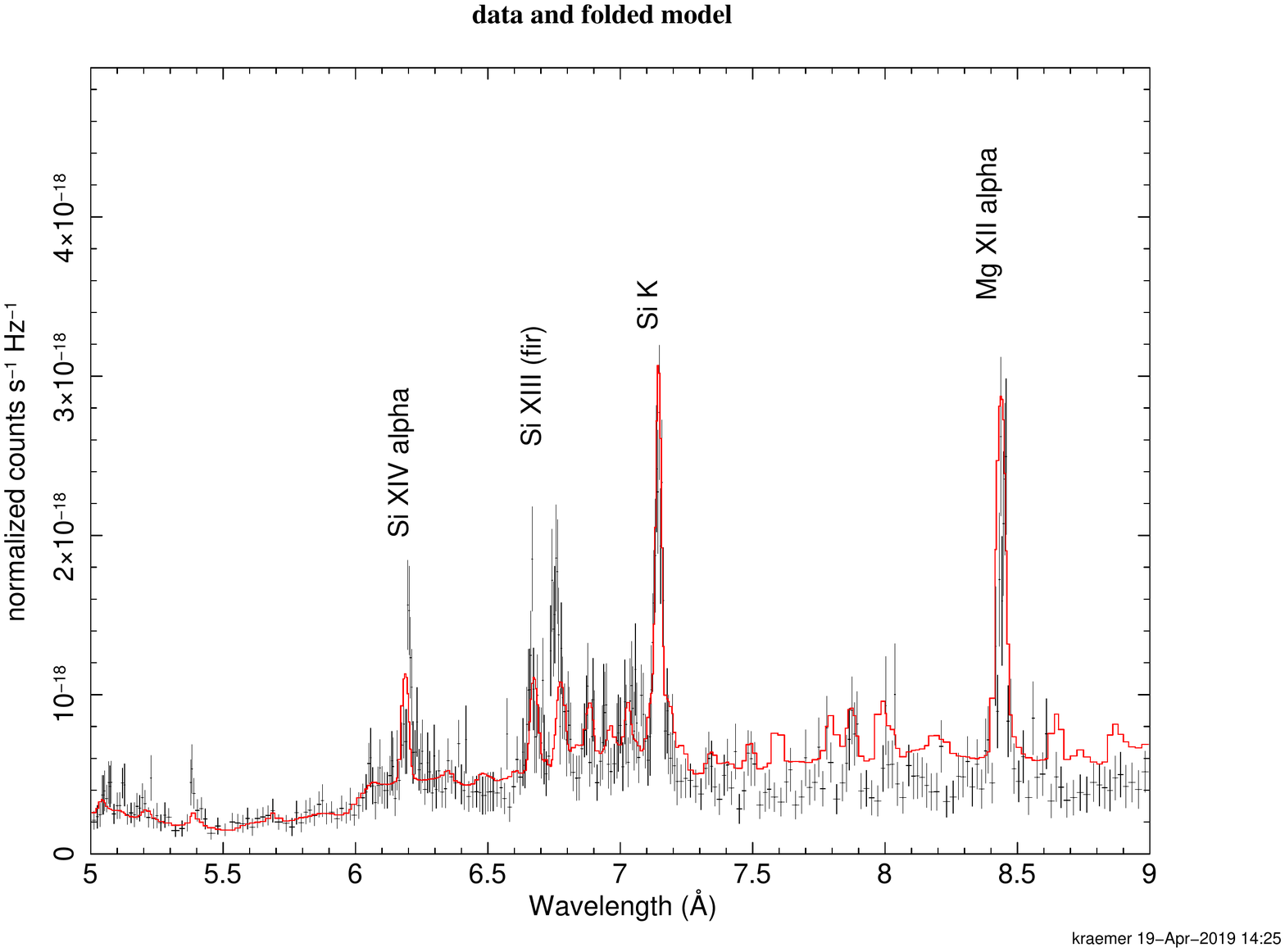}
\end{minipage}\qquad
\begin{minipage}[b]{.4\textwidth}
\includegraphics[trim= 0 40 0 0, clip, width=8.5cm]{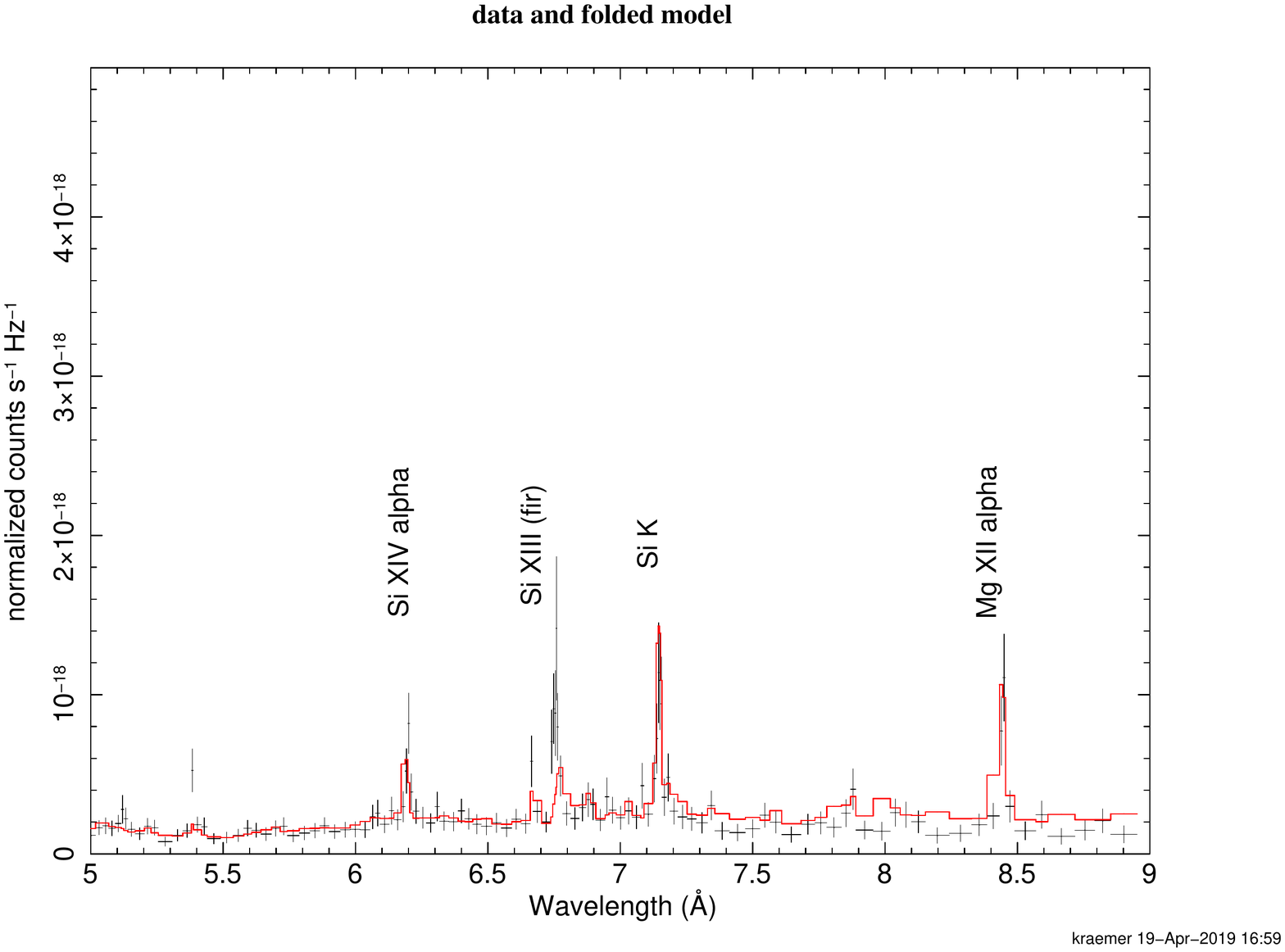}
\end{minipage}
\caption{{\it Left hand panel}: Full model, in red, plotted against the MEG data, in black, for the
range 5\AA-9\AA. Strong emission features are identified. Overall,
the model provides an adequate fit to the data, albeit with the discrepancies discussed in Section \ref{sec:InitRes}.
{\it Right hand panel}: Full model, in red, plotted against the HEG data, in black, for the
range 5\AA-9\AA.
}\label{fig:MEG59}
\end{figure*}

\begin{figure*}
\centering
\begin{minipage}[b]{.4\textwidth}
\includegraphics[trim= 20 40 0 0,clip, width=9cm]{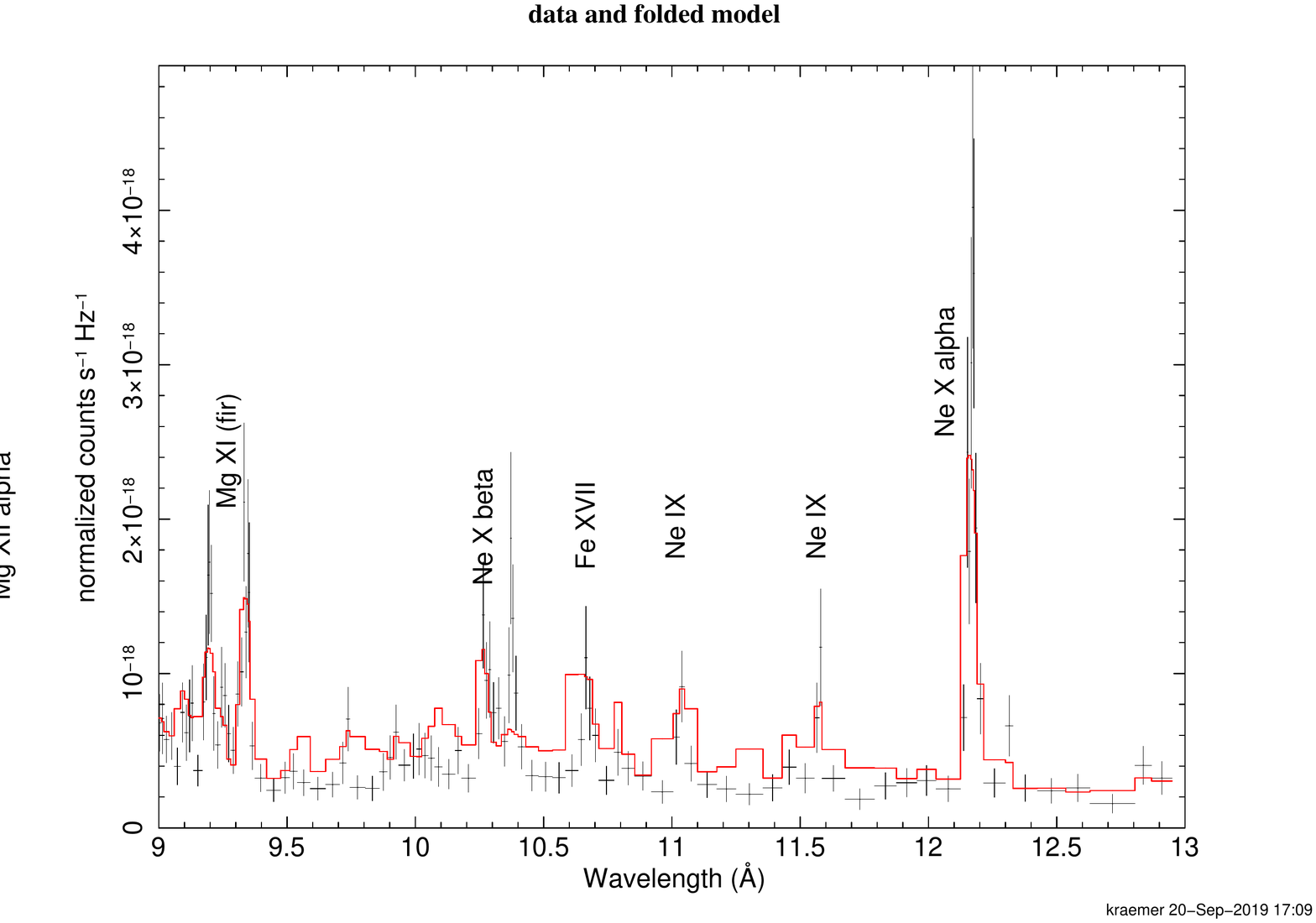}
\end{minipage}\qquad
\begin{minipage}[b]{.4\textwidth}
\includegraphics[trim= 0 40 0 0, clip, width=9cm]{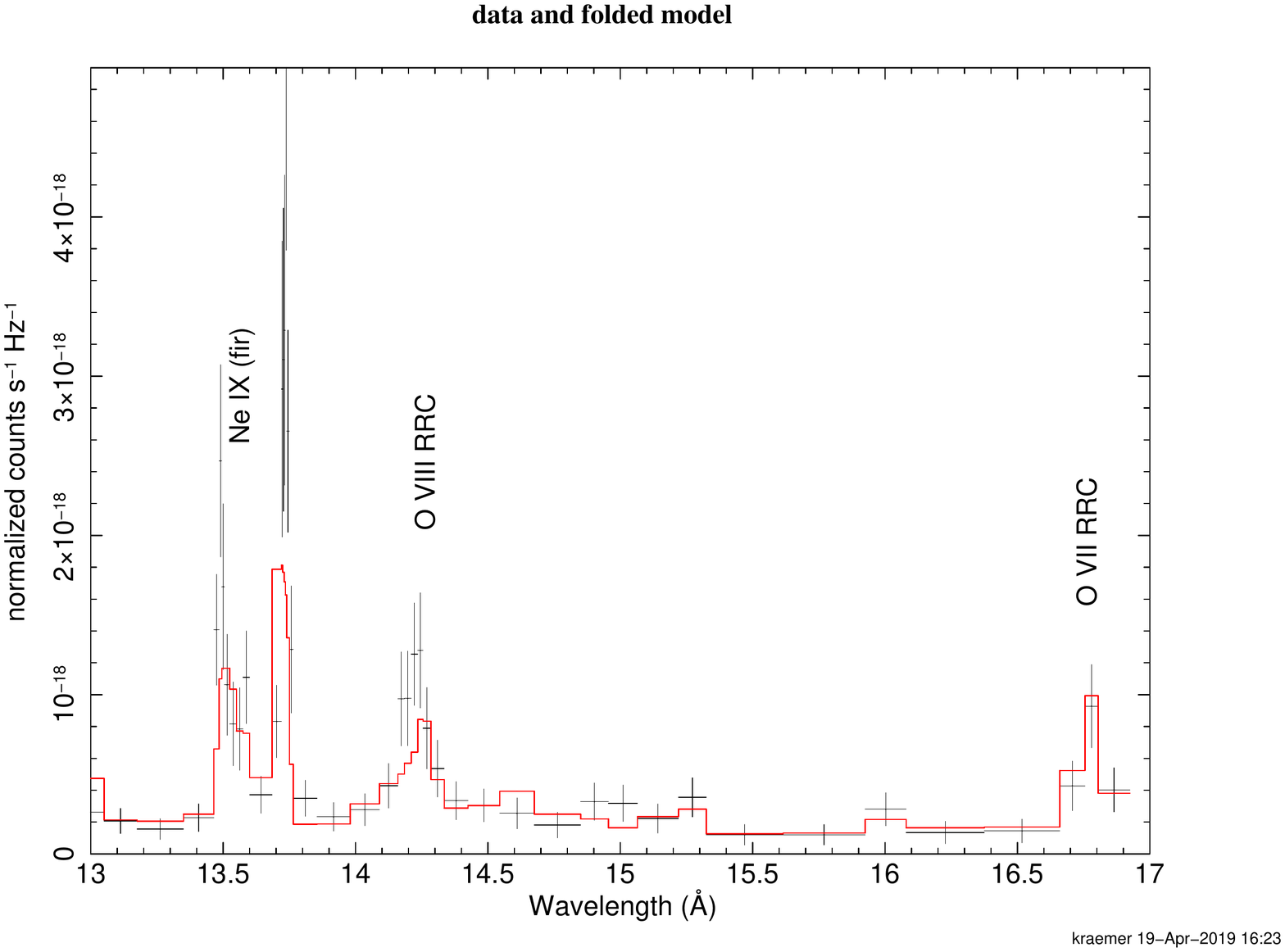}
\end{minipage}
\caption{{\it left hand panel}: Full model (red) versus MEG data (black) for the range
9\AA~- 13\AA. {\it right hand panel}: Same for the range 13\AA~- 17\AA.}
\label{fig:MEG917}
\end{figure*}

\begin{figure*}
\centering
\begin{minipage}[b]{.4\textwidth}
\includegraphics[width=8.5cm]{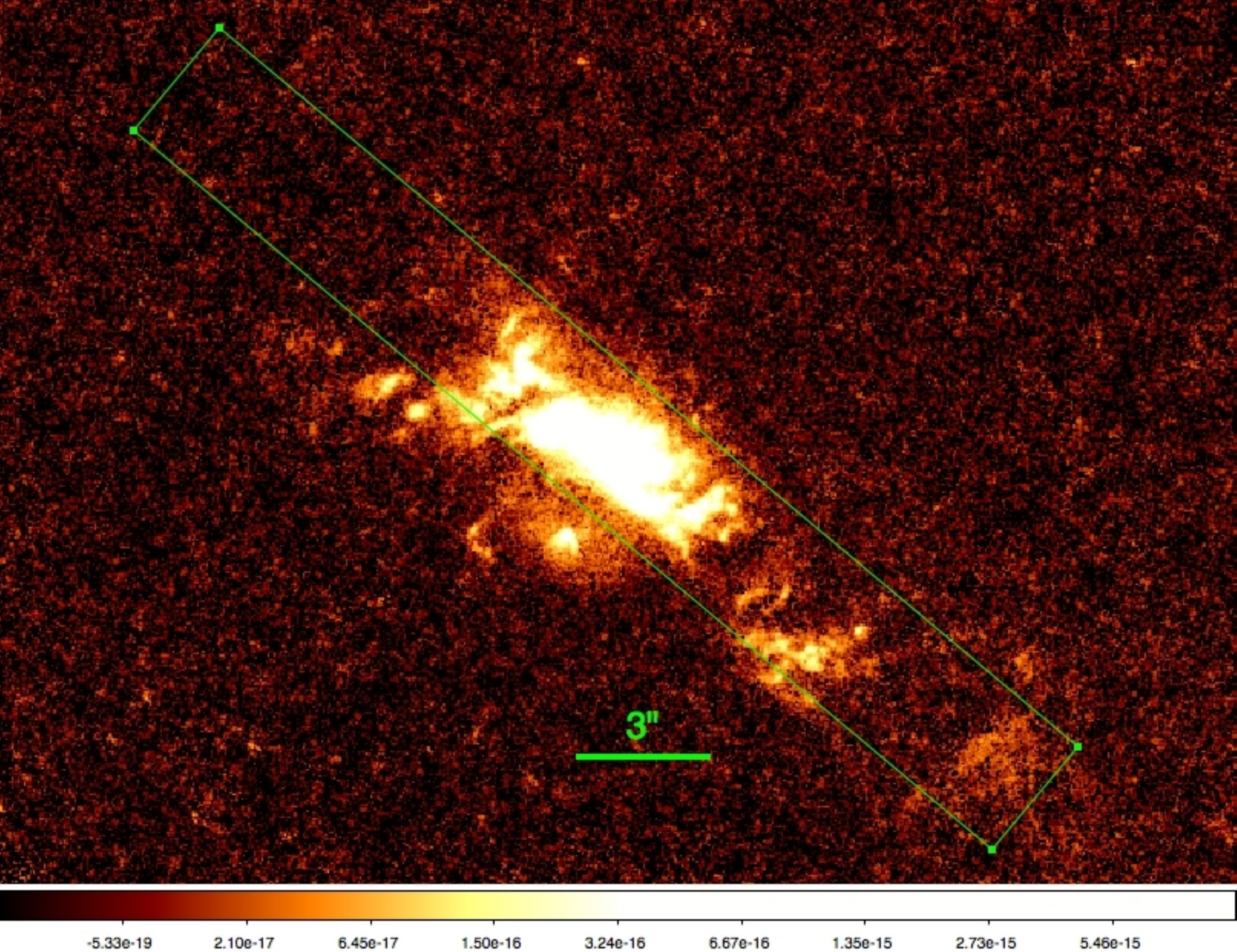}
\end{minipage}\qquad
\begin{minipage}[b]{.4\textwidth}
\includegraphics[trim= 0 230 0 0, clip, width=8.5cm]{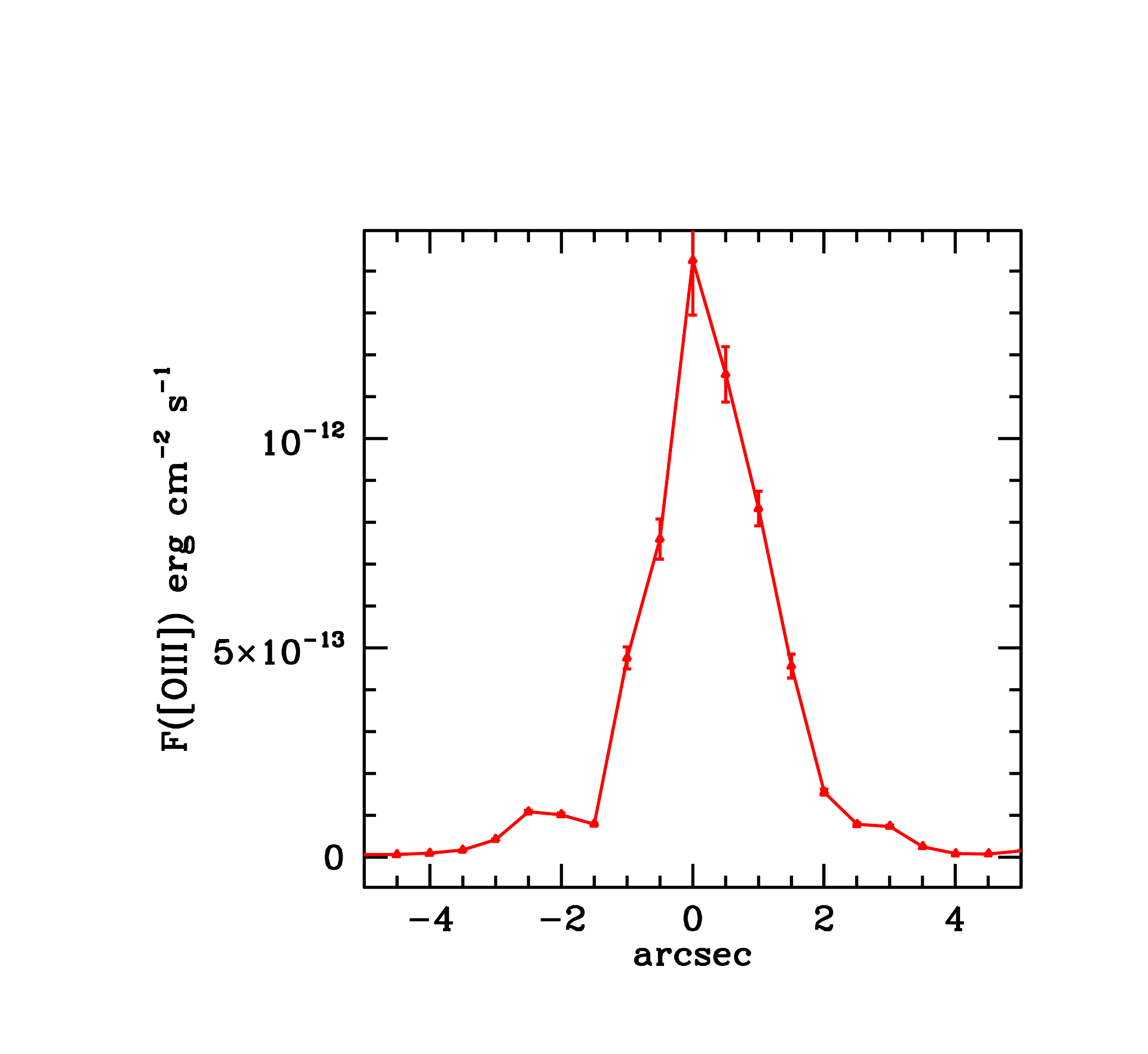}
\end{minipage}
\caption{{\it left hand panel}: The 0\arcsecpoint5 X 3\arcsecpoint0 extraction region used to obtain the Ne~IX and Ne~X profiles, along
positon angle 140\degree, superimposed
upon an archival WFPC2 [O~III] image of NGC 4151. North up, east to left. {\it right hand panel}: [O~III] fluxes measured within the same extraction bins; the error bars show the standard deviations
which were calculated by measuring
the root mean squares of fluxes in the extraction window and multiplying this number by the square root of the
number of pixels in this extraction region. Positive positions are those southwest (SW) of the nucleus.}\label{fig:XBOX}
\end{figure*}

\begin{figure*}
\centering
\begin{minipage}[b]{.4\textwidth}
\includegraphics[trim= 0 25 0 0,clip, width=6cm]{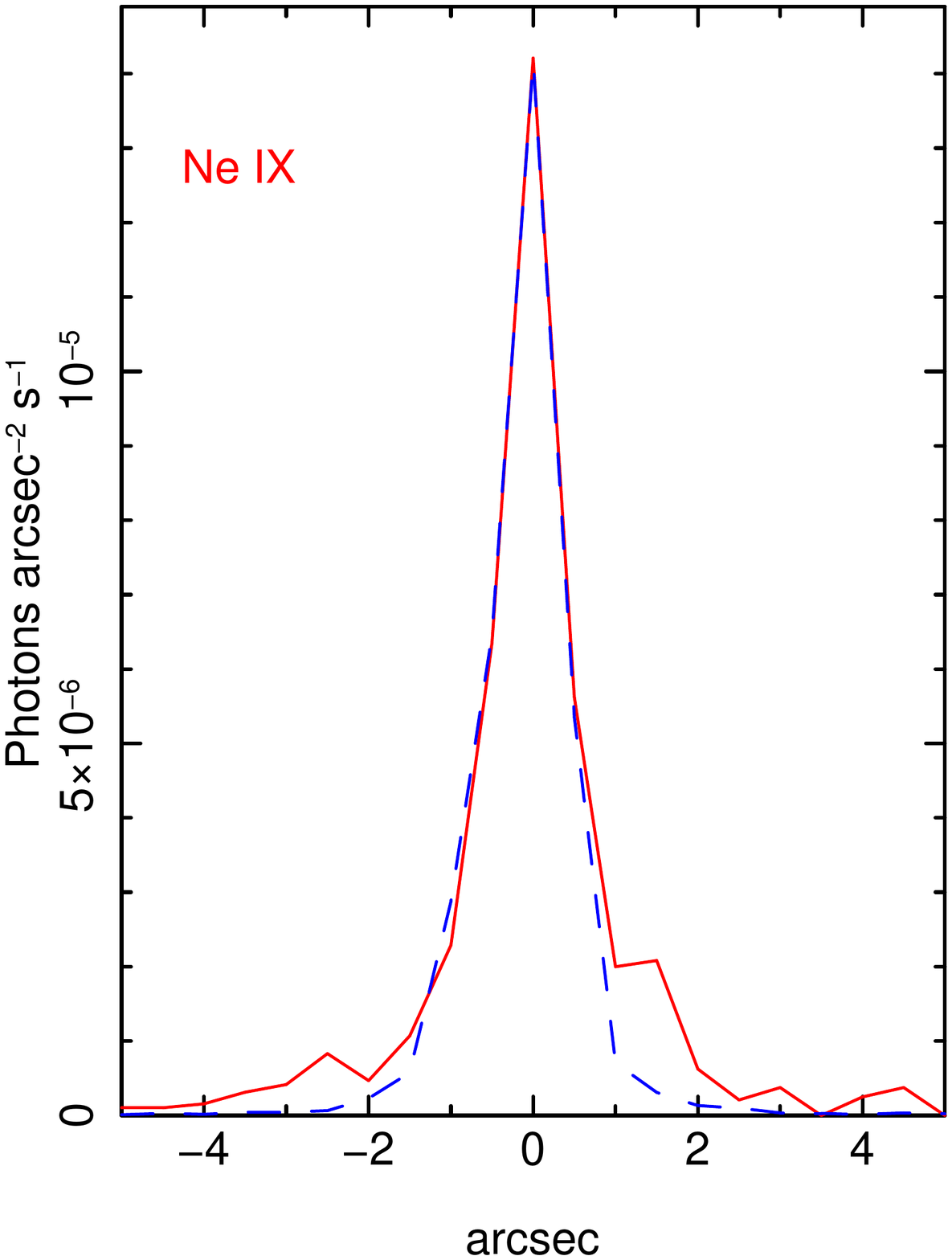}
\end{minipage}\qquad
\begin{minipage}[b]{.4\textwidth}
\includegraphics[trim = 0 40 0 0,clip,width=6cm]{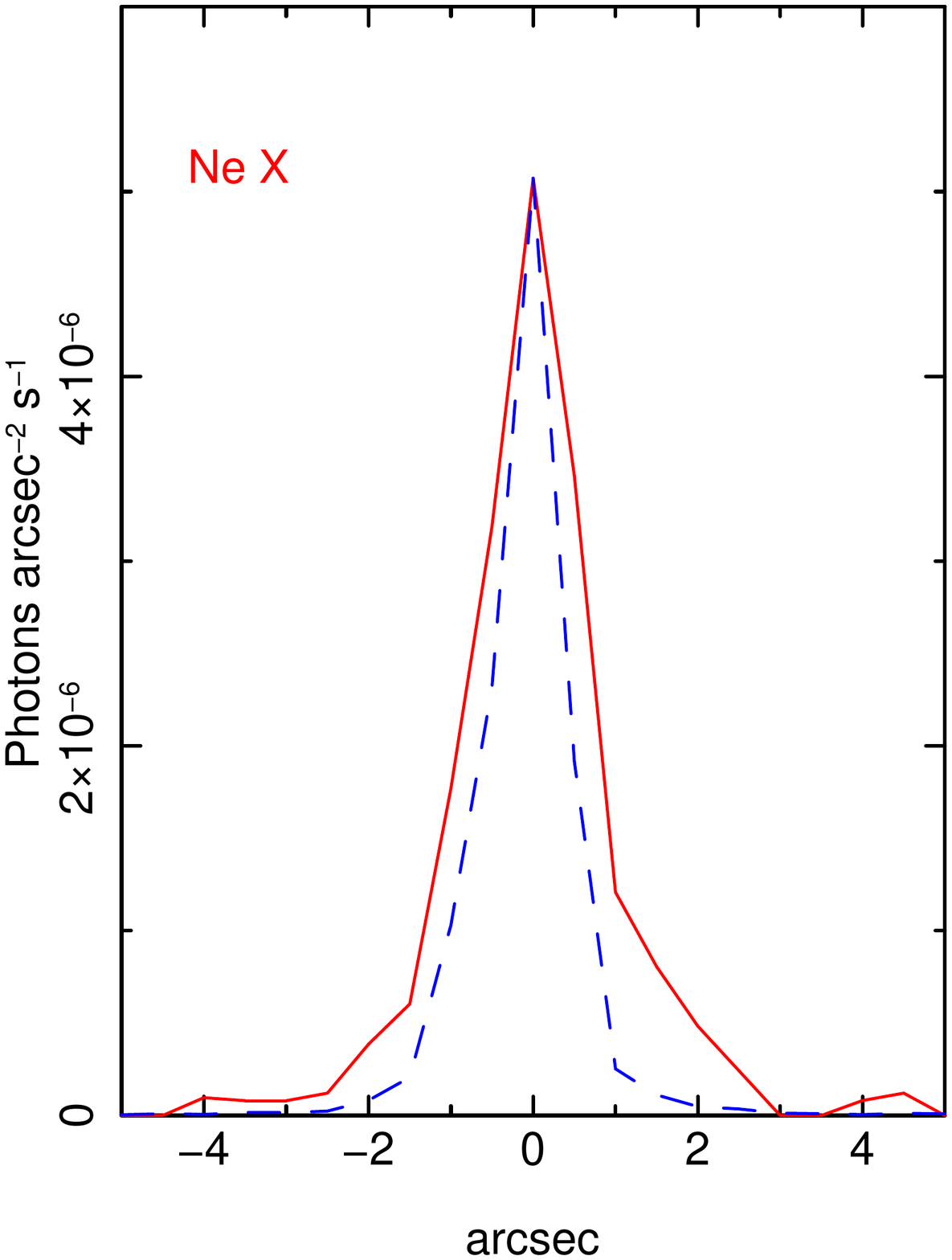}
\end{minipage}
\caption{{\it left hand panel}: Ne IX emission-line profile (in red) as a function of position. The fluxes were
measured from the zeroth order HETG spectrum. The blue dashed line is the continuum; positive positions are SW of the nucleus. {\it right hand panel}: Ne X emission-line
profile.
}\label{fig:NEIXNEX}
\end{figure*}

\begin{figure*}
\centering
\begin{minipage}[b]{.4\textwidth}
\includegraphics[trim = 0 100 0 0, clip,width=8.5cm]{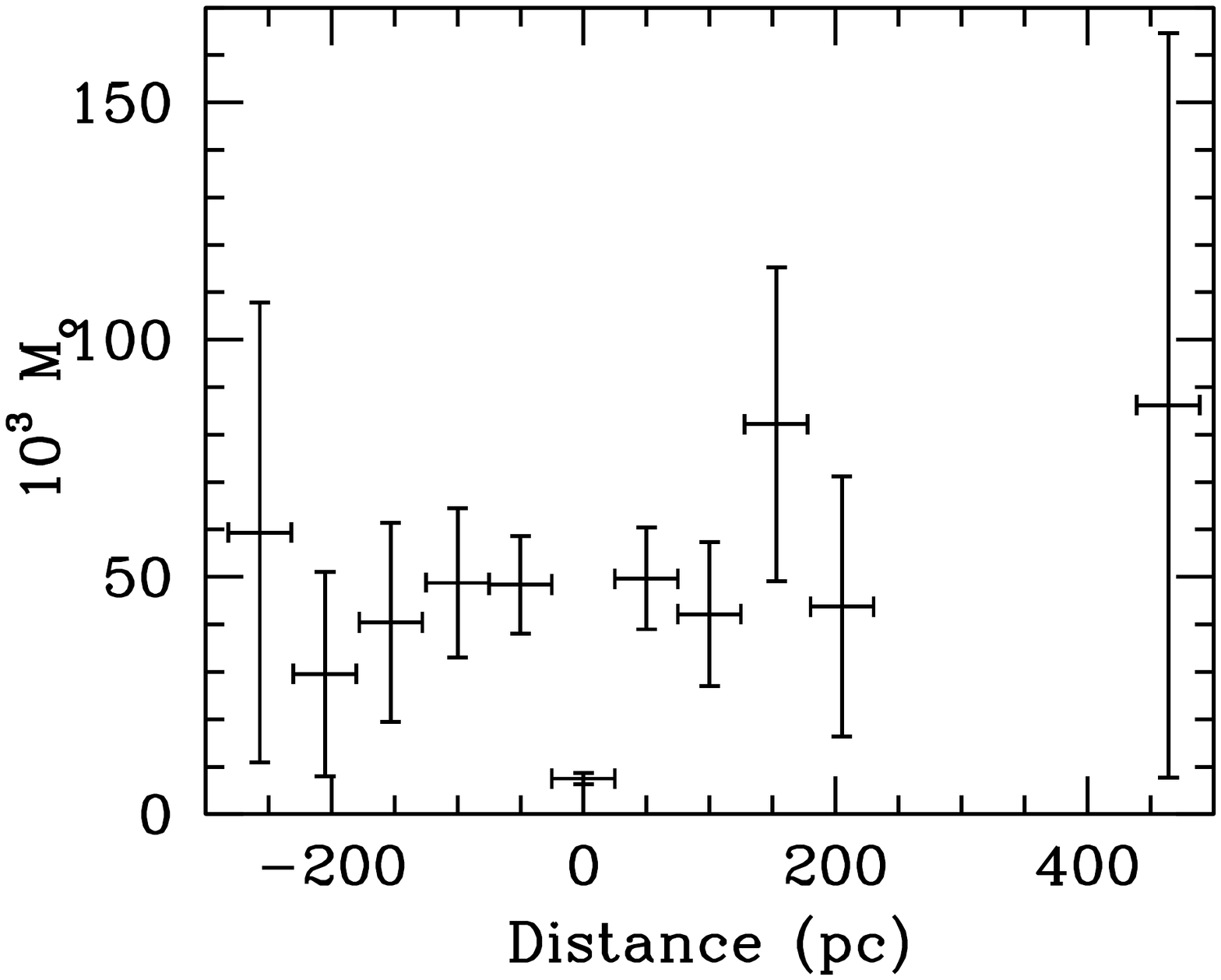}
\end{minipage}\qquad
\begin{minipage}[b]{.4\textwidth}
\includegraphics[trim = 0 100 0 0, clip, width=8.5cm]{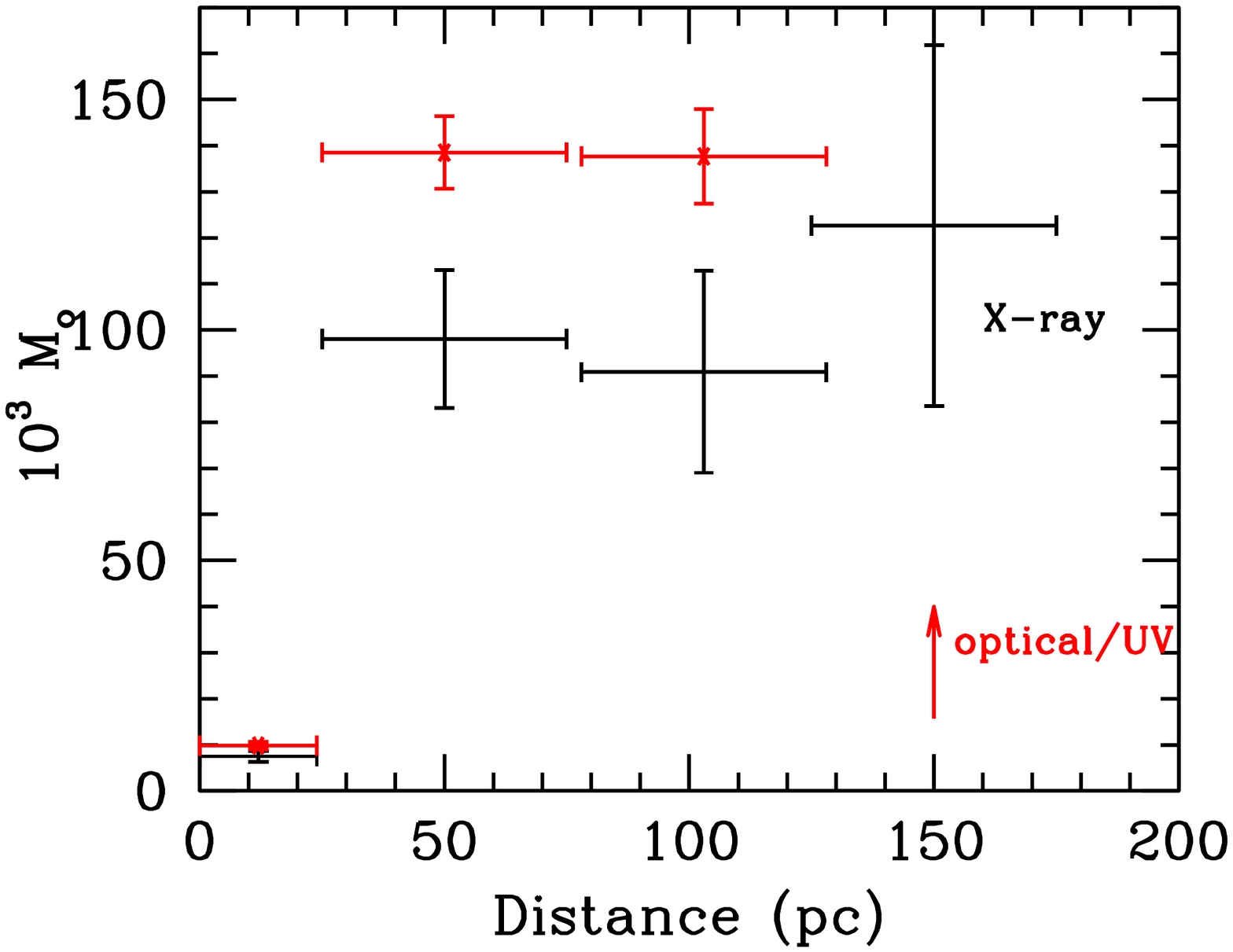}
\end{minipage}
\caption{{\it left hand panel}: Computed values of $M$, in units of 10$^{3}$ solar masses, as a function of deprojected distance, in parsecs. The positive/negative distances
correspond to point SW/NE of the nucleus, respectively. Uncertainties in $M$ value reflect photon counting statistics,
while the uncertainties in the distances correspond to the bin sizes. {\it right hand panel}:
Computed $M$ values at the same radial distances (SW and NE values summed) for
the HETG analysis (black crosses), compared to those from the STIS optical/UV analysis (red asterisks),
from \citet{crenshaw15a} and \citet{revalski18b}. The optical/UV points are summed
to correspond with the HETG extraction bin sizes. Uncertainties in the optical/UV points are those from
Crenshaw et al., and have been added in quadrature to account for the binning.}\label{fig:MMdot}
\end{figure*}

\begin{figure*}
\includegraphics[trim = 0 100 0 0,clip, width=8.5cm]{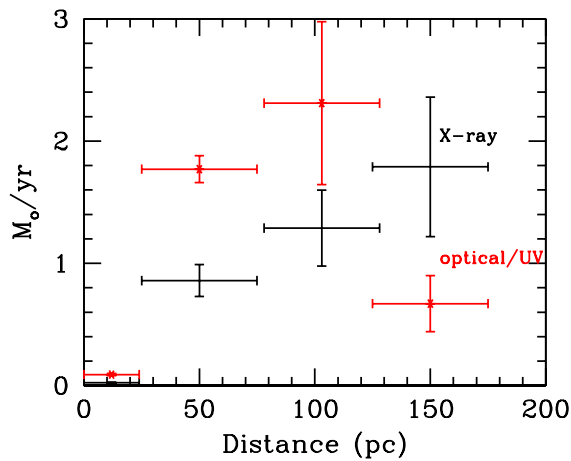}
\caption{Computed values of $\dot{M}_{out}$ as a function of distance, from the HETG analysis
(black crosses), compared to those from the STIS optical/UV analysis (red asterisks). Both both datasets, the
$\dot{M}_{out}$ values were 
computed using flux-weighted velocities within each bin from the STIS analysis. Uncertainties include both
those of the $M$ values and the flux-weighted velocities.}\label{fig:MDOTCOMP}
\end{figure*}

\begin{table*}
	\centering
	\caption{Measured Emission Features}
	\label{tab:Lines}
        \begin{threeparttable}
	\begin{tabular}{|lcccccc|} 
\hline

Ion & Ionization Potential$^{a}$ & Wavelength & Flux  & $\sigma^{b}$  & $v_{\rm off}^{c}$ & $\Delta$$\chi^{2}$$^{d}$\\  
 & (keV) & (\AA) & $(\times 10^{-6}$ cm$^{-2}$ s$^{-1})$ & (eV)  & (km s$^{-1}$) & \\
\hline
Si~XIV $\alpha$ & 2.44 & 6.202$^{+0.005}_{-0.005}$ & 10.0 $\pm$ 1.0 & 1.0 -- 5.2 &  $-$156$^{+84}_{-375}$ & -28\\
Si~XIII r & 0.52 & 6.666 & 5.4 $\pm$ 0.7 & & & -118$^{e}$   \\
Si~XIII i & & 6.707 & 2.6 $\pm$ 0.4 &  &  & \\
Si~XIII f &  & 6.758$^{+0.001}_{-0.006}$ & 8.1 $\pm$1.4 & $< 5.9$ & $-$217$^{-146}_{-463}$ &  \\
Mg~XII $\beta$ & 1.76 &  7.122 & 4.6 $\pm$ 0.7 & $< 0.4$ & & -8 \\
Mg~XII $\alpha$ & & 8.446$^{+0.005}_{-0.003}$ & 7.4 $\pm$ 0.8 &  & $-$149$^{+12}_{-255}$ & -28  \\
Ne X RRC & 1.20 & 9.17 & 12.1 $\pm$ 2.7 &  &  & -3 \\
Mg~XI r & 0.37 &  9.193 & 7.4 $\pm$ 1.1 &  & & -25$^{e}$ \\
Mg~XI i & & 9.258 & 3.1 $\pm$ 0.9 &  &  & \\
Mg~XI f & &  9.337$^{+0.006}_{-0.007}$ & 10.0 $\pm$ 1.5 &  & $-$244$^{-37}_{-452}$  &  \\
Ne~IX RRC & 0.24 & 10.37 & 5.6 $\pm$ 1.0 &  & & -30 \\
Ne~X $\alpha$ & & 12.169$^{+0.003}_{-0.009}$ & 29.5 $\pm$ 3.0 &  & $-$284$^{-66}_{-353}$  & -40\\
Ne~IX r & & 13.50 & 25.2 $\pm$ 6.0 &  & & -45$^{e,f}$ \\
Ne~IX i & & 13.60 & 23.1 $\pm$ 6.0 & & & \\
Ne~IX f & & 13.73 & 72.5 $\pm$ 7.0 & $<1.0$ & $-$295 (fixed) & \\
O~VIII RRC & 0.74 & 14.25 & 42.0 $\pm$ 6.0 &  & &  \\
O~VII RRC & 0.14  & 16.82 & 56.0 $\pm$ 13.0 &  & & \\
O~VIII $\alpha$ & & 19.05$^{+0.09}_{-0.13}$ & 104.0 $\pm$ 19.0 & $< 1.0$ & $+$231$^{+1730}_{-1852}$ & \\
O~VII r & & 21.71  & 72.3 $\pm$ 43.0 & & & \\
O~VII f & & 22.19$^{+0.03}_{-0.01}$ & 378.0 $\pm$ 85.0 & 0.4 -- 2.0 & $-$236$^{+199}_{-412}$  &\\     
\hline 
\end{tabular}
\begin{tablenotes}
\item[$^{a}$] Ionization Potentials from \citet{cox00a}.
\item[$^{b}$] Line width, in eV, derived from a Gaussian fit to the line. When blank,
the widths were fixed at 2 eV, for Gaussian lines, and $kT_{e}=$ 5 eV for RRCs,
where $k$ is Boltmann's constant and $T_{e}$ is the electron temperature.
\item[$^{c}$] Blank fields indicate no constraints were obtained.
\item[$^{d}$] Significance of feature returned by the sliding Gaussian test (Section \ref{sec:Inputs}; Figure \ref{fig:Slide}). Features at 
$\lambda > 15$ \AA~were not tested, due to the weakness of the continuum.
\item[$^{e}$] Combined for the full triplet
\item[$^{f}$] Combined with O~VIII RRC.
\end{tablenotes}
\end{threeparttable}
\end{table*}

\begin{table*}
	\centering
	\caption{Fit Parameters for the HETG data$^{a}$}
	\label{tab:Fits}
	\begin{threeparttable}
	\begin{tabular}{|lcccc|} 
\hline
ATABLES & & & &\\
\hline
Zone & $log N_{\rm H}$  & $Log U$  & Norm$^{b}$ &  Velocity \\  
 &    ${\rm cm^{-2}}$  &  & & km/s \\
\hline
HIGHION & 22.5  & 1.0    & 8.0e-17 &  $-369\pm37$   \\
MEDION  & 22.54 $\pm$ 0.12   & 0.19$\pm$ 0.07 & 3.0($\pm 1.2$)e-16 &  76$\pm45$ \\
LOWION  & 23.0 & - 0.50 & 2.9e-16 & 86  \\
\hline 
\end{tabular}
\begin{tabular}{|lccc|} 
\hline
MTABLES & & & \\
\hline
Zone & $log N_{\rm H}$  & $Log U$  &  Velocity \\  
 &    ${\rm cm^{-2}}$  &   & km/s \\
\hline
1 & 22.7 $\pm$ 0.02  & $-1.8$    &  0   \\
2  & 22.8    & $-1.8$  &  0 \\
\hline 
\end{tabular}
\begin{tabular}{|lcccc|} 
\hline
REFLECTION$^{c}$ & & & & \\
\hline
model &  $logXi$ & Incl. & Norm$^{b}$ & Velocity\\  
 &  & deg. && km/s \\
\hline
xillverEc & 1.06 $\pm$0.05 & 49.45 $\pm$26.9 & 1.18($\pm 0.10$)e-04  &  696   \\
\hline 
\end{tabular}
\begin{tablenotes}
\item[$^{a}$] All parameters frozen, unless otherwise indicated.
\item[$^{b}$] Normalization of the model components during the {\sc XSPEC} fitting. 
\item[$^{c}$] For xillver model, we assumed $\Gamma=1.5$, solar Fe abundance, and an energy cutoff of 500 eV.
\end{tablenotes}
\end{threeparttable}
\end{table*}

\begin{table*}
	\centering
	\caption{Cloudy Model Parameters$^{a}$}
	\label{tab:Inputs}
	\begin{threeparttable}
	\begin{tabular}{|lcccccc|} 
\hline
Model & log($n_{\rm H})$ & distance & $A$ & $C_{f}$ & FM$^{b}$ & T$_{e}$ \\
 & (cm$^{-3})$ & (cm) & (cm$^{2}$) & & &(K) \\ 
 \hline
HIGHION & 4.10 & 2.5e18 & 3.2e36 & 0.040 & 1.44/1.22 & 2.94e06\\
MEDION & 3.37 & 1.5e19 & 6.4e37 & 0.023 & 10.2/3.54 & 1.69e05\\
LOWION & 3.60 & 2.5e19 & 3.6e37 & 0.004 & 53.8/6.19 & 1.91e04\\
\hline 
\end{tabular}
\begin{tablenotes}
\item[$^{a}$] Assuming $\Delta r/r < 1$, see text.
\item[$^{b}$] Predicted Force Multiplier (FM) values are given for initial/final zone. 
\end{tablenotes}
\end{threeparttable}
\end{table*}

\begin{table*}
	\centering
	\caption{Comparison of Measured and Predicted Emission Line Luminosities$^{a}$}
	\label{tab:FinalFit}
	\begin{threeparttable}
	\begin{tabular}{|lcccccc|} 
\hline
Line & Measured Luminosity  & HIGHION  & MEDION & LOWION & Total$^{b}$ & data/model\\  \hline
Si~XIV $\alpha$ & 7.40$\pm$0.74 & 4.49 & 0.84 & & 5.33 & 1.39 \\
Si~XIII r & 3.74$\pm$0.48 & 1.55 & 2.21 &  & 3.77 & 0.99\\
Si~XIII i & 1.79$\pm$0.28 & 0.18 & 0.95 & & 1.13 & 1.59 \\
Si~XIII f & 5.50$\pm$0.95 & 0.41 & 2.00 & & 2.40 & 2.29\\
Mg~XII $\beta$ & 2.96$\pm$0.45 & 1.04 & 0.67 & & 1.70 & 1.74\\
Mg~XII $\alpha$ & 4.83$\pm$0.52 & 3.29 & 2.32 & & 5.61 & 0.86\\
Ne X RRC & 6.06$\pm$1.35 & 3.34 & 2.33 & & 5.67 & 1.07\\
Mg~XI r & 3.70$\pm$0.55 & 0.56 & 3.44 & 0.07 & 4.07 & 0.91 \\
Mg~XI i & 1.53$\pm$0.44 & 0.04 & 2.25 & & 2.29 & 0.67 \\
Mg~XI f & 4.73$\pm$0.71 & 0.12 & 6.03 & 0.02 & 6.17 & 0.77 \\
Ne~IX RRC & 2.46$\pm$0.44 & 0.07 & 2.91 & 0.08 & 3.07 & 0.80\\
Ne~X $\alpha$ & 10.1$\pm$1.03 & 4.07 & 9.36 & 0.12 & 13.5 & 0.75\\
Ne~IX r & 7.87$\pm$1.87 & 0.41 & 6.10 & 0.45 & 6.96 & 1.13\\
Ne~IX i & 7.82$\pm$2.03 & 0.03 & 5.93 & 0.22 & 6.18 & 1.27\\
Ne~IX f & 18.7$\pm$1.87 & 0.09 & 18.5 & 0.75 & 19.3 & 0.97\\
O~VIII RRC & 13.5$\pm$1.93 & 8.92 & 14.9 & 0.27 & 24.1 & 0.56\\
O~VII RRC & 15.3$\pm$3.55 & 0.05 & 8.44 & 1.03 & 9.52 & 1.61 \\
O~VIII $\alpha$ & 38.4$\pm$7.03 & 7.86 & 35.2 & 1.87 & 44.9 & 0.86 \\
O~VII r & 9.62$\pm$5.72 & 0.44 & 9.36 & 2.47 & 12.3 & 0.78 \\
O~VII f & 78.4$\pm$17.6 & 0.12 & 47.0 & 17.5 & 64.6 & 1.21 \\     
\hline 
\end{tabular}
\begin{tablenotes}
\item[$^{a}$] In units of 10$^{38}$ erg s$^{-1}$.
\item[$^{b}$] To determine the line luminosities, the predicted line fluxes were multiplied by the
emitting areas listed in Table 3.  
\end{tablenotes}
\end{threeparttable}
\end{table*}

\begin{table*}
	\centering
	\caption{Cloudy Model Parameters for Spatially Resolved Ne~IX and Ne~X Emission}
	\label{tab:SpatRes}
	\begin{threeparttable}
	\begin{tabular}{|lccccl|} 
\hline
Model & distance & log$U$& log($n_{\rm H}$)$^{a}$ & log($N_{\rm H}$)$^{b}$ & FM$^{c}$\\ 
 &  & & cm$^{-3}$ & cm$^{-2}$ & \\
 \hline
HIGHION & 12 pc & 0.59 & 2.17 & 21.74 & 2.74/1.95 \\
 & 50 pc & 0.37& 1.15 & 21.34 & 4.86/3.26\\
 & 100 pc & 0.26 & 0.66 & 20.85 & 6.68/5.37\\
 & 153 pc & 0.20 & 0.35 & 20.54 & 10.0/7.22\\
 & 205 pc & 0.16 & 0.14 & 20.33 & 11.9/8.78\\
 & 257 pc & 0.12 & -0.02 & 20.17 & 45.0/17.2\\
 & 309 pc & 0.09 & -0.15 & 20.04 & 48.2/19.1\\
 & 360 pc & 0.07 & -0.26 & 19.93 & 50.4/20.8\\
 & 412 pc & 0.05 & -0.36 & 19.83 & 52.7/22.6\\
 & 464 pc & 0.03 & -0.44 & 19.67 & 18.9/14.9$^{d}$\\
\hline
MEDION &  12 pc & 0.05 & 2.71 & 22.28 & 16.7/4.44  \\
 & 50 pc & -0.17 & 1.69  & 21.88 & 26.0/6.92\\
 & 100 pc & -0.27 & 1.19 & 21.38 & 31.8/9.96\\
 & 153 pc & -0.34 & 0.89 & 21.06 & 37.4/12.7\\
 & 205 pc & -0.38 & 0.68 & 20.81 & 41.1/14.9\\
 & 257 pc & -0.42 & 0.52 & 20.71 & 45.0/17.2\\
 & 309 pc & -0.45 & 0.39 & 20.58 & 48.2/19.1\\
 & 360 pc & -0.47 & 0.28 & 20.27& 50.4/20.8\\
 & 412 pc & -0.49 & 0.18 & 20.37 & 52.7/22.6\\
 & 464 pc & -0.50 & 0.09 & 20.28 & 53.9/24.0\\ 
\hline 
\end{tabular}
\begin{tablenotes}
\item[$^{a}$] Based on density law $n_{\rm H} \propto r^{-1.65}$. 
\item[$^{b}$] Assuming $\Delta r/r < 1$ or $\Delta r$ fixed at 50 pc; for a detailed explanation see Section \ref{sec:MMdot2}. 
\item[$^{c}$] Predicted Force Multiplier (FM) values are given for initial/final zone. 
\item[$^{d}$] The drop in FM is due to the small value for $N_{\rm H}$, which results
in a higher average ionisation compared to the model predictions for 412 pc. Cloudy calculates the radiation transfer and physical
conditions over zones of finite size. Typically these are scaled in units of optical depth, rather than physical depth.
So, if the zone is small in depth, the average ionization can be higher than a similar zone that penetrates deeper into the gas.
Since the depth of the model at 412 pc is small compared to the other regions, it is of higher average ionisation. Hence the FM,
which depends on the ionisation state of the gas, is lower than it would be for a model with the same $U$, but with a greater $N_{\rm H}$.
\end{tablenotes}
\end{threeparttable}
\end{table*}




\bibliographystyle{mnras}
\bibliography{kraemersb} 



\appendix

\counterwithin{figure}{section}
\counterwithin{table}{section}

\section{Appendix A}\label{sec:App}

As noted in Section \ref{sec:Spectra}, our analysis was done with the data binned to a minimum of 10 counts per spectral bin.
In order to determine if our binning affected the results, we 
performed a test for which we rebinned the data to a minimum of 20 counts per spectral bin.
We refitted the combined HEG and MEG spectra, allowing key parameters for the {\sc MTABLES}, {\sc ATABLES},
and the reflection model, {\sc xillverEC}, to vary. The values that varied, compared to those for the
10 counts per bin fitting, are listed in Table \ref{tab:Fits20}.
Other than some shifts in velocity, which would be expected as emission features spread out with the
coarser binning, the results are essentially unaffected.

\begin{figure*}
\centering
\begin{minipage}[b]{.40\textwidth}
\includegraphics[trim= 0 40 0 0,clip, width=8cm]{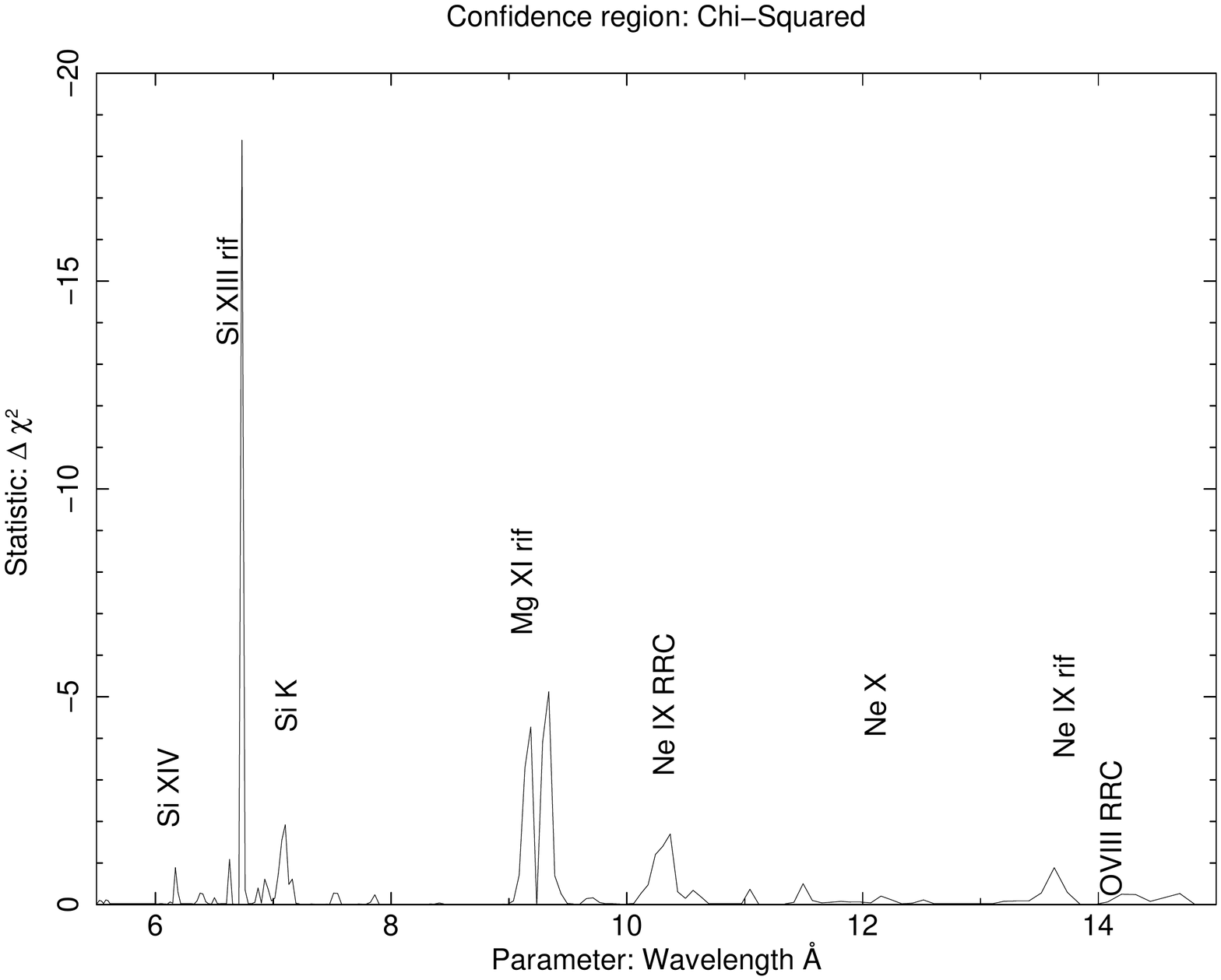}
\end{minipage}\qquad
\begin{minipage}[b]{.40\textwidth}
\includegraphics[trim = 0 40 0 0,clip,width=8cm]{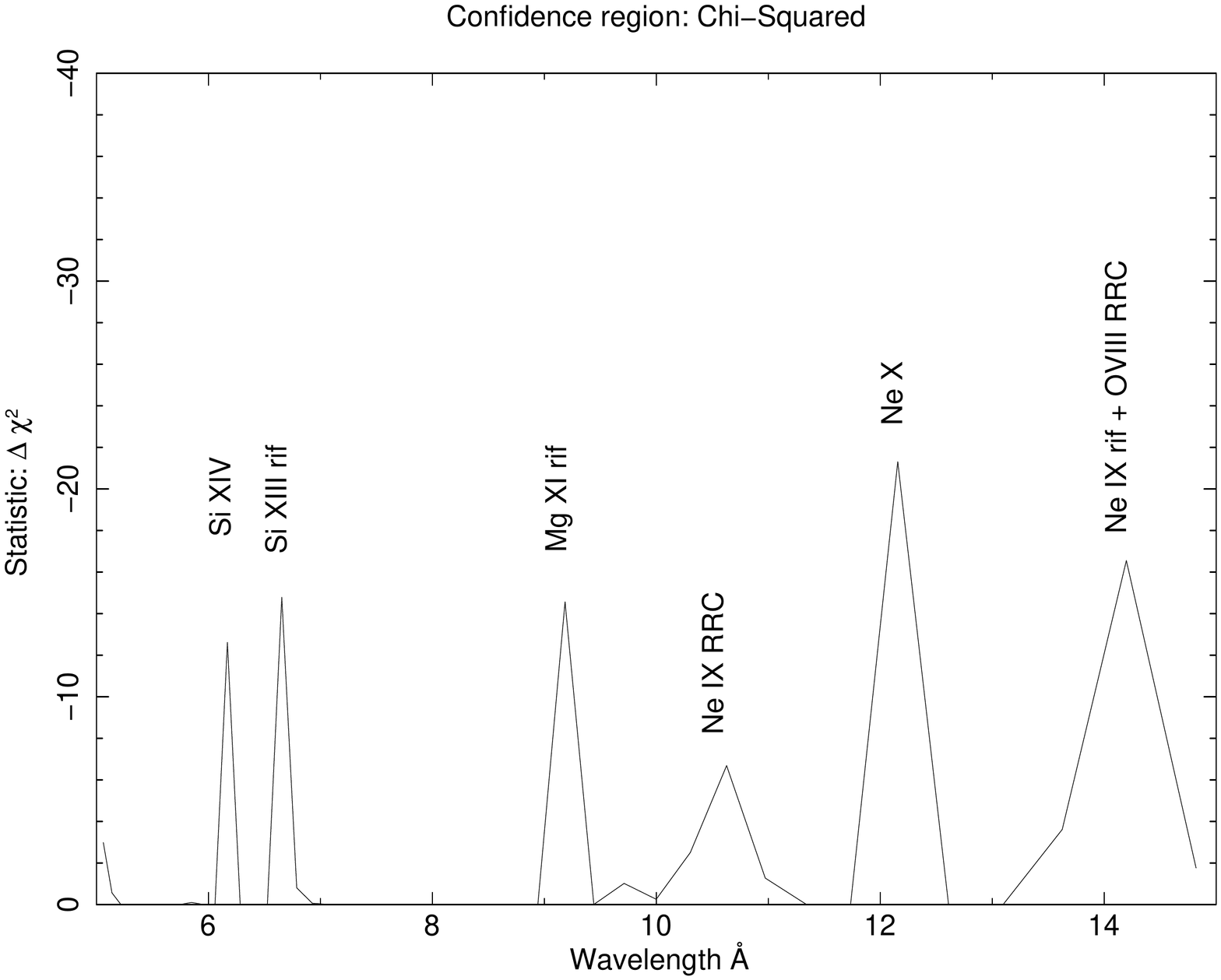}
\end{minipage}
\caption{Sliding Gaussians tests, over the range 5.5\AA-15\AA, for the HEG (left hand panel) and MEG (right hand panel)
spectra, for a minimum of 20 counts per spectral bin. The test was performed on the spectra with the inclusion of the absorbed power-law and reflection
components, refitted for 20 counts/bin (see Table\ref{tab:Fits20}). The detected emission features are labeled.}\label{fig:Slide20}
\end{figure*}

To test whether the significance of the emission features had changed, we reran the sliding Gaussian test
described in Section \ref{sec:Inputs}. The results are shown in Figure \ref{fig:Slide20}.
As anticipated (see Section \ref{sec:Spectra}), several weak features are weaker or undetected
in the 20 count binned spectra. Specfically, in the HEG, Ne~X~$\alpha$ and the O~VIII RRC are weaker,
while in the MEG, the Si~K, Mg~XI~$\beta$, and Ne~IX~$\beta$ are undetected, although the weakness of the Si~K is due to the subtraction
of the {\sc xillverEC} component. However none of these features were used to constrain the {\sc ATABLES} or photoionisation 
models. 

Two other differences highlight problems with the coarser binning. First, the Si~XIII triplet is significantly
stronger in the 20 counts per bin test. This is due to the inclusion of the unmodeled emission, discussed in
Section \ref{sec:InitRes}, in the same bin as the Si~XIII lines. The more significant difference is the absence
of the Mg~XII~$\alpha$ line in either the HEG or MEG tests. The line is clearly present in the spectra, as shown on Figure \ref{fig:FullSpec},
however, it is at a minimum point in the continuum, hence the flux at this wavelength is dominated by the line emission.
In the 20 count binning, the line flux is spread out in the bin, which makes it appear as continuum emission in the
sliding Gaussian test. Nevertheless, Si~XIV~$\alpha$ and Ne~X~$\alpha$, the line diagnostics for the component in which the Mg~XII forms, HIGHION,
are detected by the sliding Gaussian. Furthermore, the photoionisation models provide a good fit for Mg~XII~$\alpha$,
as shown in Table \ref{tab:FinalFit} and Figure \ref{fig:MEG59}. Therefore, we believe that the
detection of Mg~XII $\alpha$ in the 10 counts per bin analysis is reliable. 

In summary, the choice of a minimum of 10 counts per bin did not cause significant unreliability
in our analysis. Furthermore, the 20 count binning introduced some problems in line detections, due to 
the smearing out of features or the inclusion of surrounding continuum in the same spectral bins
as the lines. 

\begin{table*}
	\centering
	\caption{Comparison of Fit Parameters for the HETG data}
	\label{tab:Fits20}
	\begin{threeparttable}
	\begin{tabular}{|lc|} 
\hline
ATABLES & \\
\hline
Zone &  Velocity \\  
 & km/s \\
\hline
HIGHION 10 cts/bin  &  $-369\pm37$   \\
HIGHION 20 cts/bin & $-226\pm43$ \\
MEDION 10 cts/bin &76$\pm45$ \\
NEDION 20 cts/bin & 4$\pm43$ \\
\hline 
\end{tabular}
\begin{tabular}{|lcc|} 
\hline
MTABLES & & \\
\hline
Zone & $log N_{\rm H}$  & Norm$^{a}$ \\  
 &    ${\rm cm^{-2}}$  &  \\
\hline
Zone 1, 10 cts/bin  & 22.7 $\pm$ 0.02  & 5.35($\pm 0.22$)e-02 \\
2one 1, 20 cts/bin   & 22.7 $\pm$ 0.02 & 5.36($\pm 0.20$)e-02  \\
\hline 
\end{tabular}
\begin{tabular}{|lcc|} 
\hline
REFLECTION & &\\
\hline
model &  $logXi$ & Norm$^{a}$ \\  
\hline
xillverEc, 10 cts/bin & 1.06 $\pm$0.05 & 1.18($\pm 0.10$)e-04   \\
xillverEc, 20 cts/bin & 1.06 $\pm$0.06 & 1.18($\pm 0.11$)e-04 \\
\hline 
\end{tabular}
\begin{tablenotes}
\item[$^{a}$] Normalization of the model components during the {\sc XSPEC} fitting. 
\end{tablenotes}
\end{threeparttable}
\end{table*}

\bsp	
\label{lastpage}
\end{document}